\documentclass[12pt]{article}
\usepackage{newtxtext,newtxmath}
\usepackage{graphicx}
\usepackage[letterpaper,margin=1in]{geometry}
\usepackage{bm}
\usepackage[colorlinks=true,allcolors=blue]{hyperref}
\usepackage{booktabs}
\usepackage{tikz}
\usepackage[caption=false]{subfig} 
\usepackage{verbatim}
\usepackage{array}     
\usepackage{enumitem}

\usetikzlibrary{arrows.meta,positioning,matrix,fit,backgrounds,calc,decorations.pathreplacing}

\renewenvironment{abstract}
	{\quotation}
	{\endquotation}

\date{}

\makeatletter
\renewcommand{\fnum@figure}{\textbf{Figure \thefigure}}
\renewcommand{\fnum@table}{\textbf{Table \thetable}}
\makeatother

\usepackage{scicite}

\usepackage{url}

\def\scititle{Coherence as Thermodynamic Organization:\\ Toward a Non-Equilibrium Turbulence Theory
}
\title{\bfseries \boldmath \scititle}

\author{
	Sharath S. Girimaji$^{1\ast\dagger}$\\
	\small$^{1}$Ocean Engineering Department, Texas A\&M University, College Station, TX \& 77843, USA.\\
	\small$^\ast$Corresponding author. Email: girimaji@tamu.edu
}

\begin{document}

\maketitle

\begin{abstract} \bfseries \boldmath
  
Since the foundational studies in the late nineteenth century, fluid turbulence has stood as a profound, unsolved challenge in classical physics. Much of this enduring difficulty stems from non-equilibrium turbulence, where the lack of a unifying physical framework for macroscopic coherent structures has hampered predictive flow modeling.
Here, we establish a foundational bridge between non-equilibrium statistical physics and turbulent coherent structures through the renormalized Navier-Stokes equations. We demonstrate that all forms of turbulent coherence are fundamentally universal thermodynamic responses mandated by macroscopic energy throughput imbalances. Depending on topological access to bifurcations, these formations manifest either as transient adjustments (analogous to Kubo's near-equilibrium fluctuations) or as autonomous, transformative states (mirroring Prigogine's far-from-equilibrium dissipative structures).
By introducing a computable, effective thermodynamic order parameter ($\Pi$), this paradigm establishes a rigorous foundation for non-equilibrium theory, enabling unequivocal identification of necessary flow resolution in continuously driven, dissipative continuum systems.
\end{abstract}


Richard Feynman famously singled out fluid turbulence as the most important unsolved problem of classical physics \cite{feynman1964lectures}. In externally forced, open turbulent systems---such as wakes, jets, and boundary layers---the intractability stems largely from non-equilibrium effects manifesting via coherent structures. While classical hydrodynamic theories \cite{Kolmogorov1941, Kolmogorov1962, Batchelor1953, MoninYaglom1975} rooted in equilibrium reasoning have proven indispensable for quantifying chaotic, structureless turbulence, they systematically fail to predict or explain the fundamental features of these persistent, organized motions. From a statistical mechanics viewpoint, this limitation arises because the onset of macroscopic coherent structures signals a fundamental departure from the equilibrium stochastic regime, marking the point where baseline assumptions---principally the decorrelation of microscopic states (\textit{Stosszahlansatz})---lose their validity, primarily through the emergence of non-Markovian memory effects \cite{Girimaji2024_NJP}. A rigorous framework for coherence is therefore not merely a refinement of existing models; it is a necessary step toward a fundamental statistical theory for non-equilibrium turbulence.

The notion of coherent structures has played a formative role in turbulence research for more than half a century \cite{Lumley1981_CoherentStructures, Hussain1986, Sirovich1987_CoherentStructures}. These large-scale formations arise where continuous energy input from the mean flow drives profound departures from equilibrium \cite{Hussain1986, Townsend1956, Lumley1967}. In these conditions, macroscopic coherence and transport co-emerge: the organized motions significantly augment the baseline stochastic (background turbulence) transport. Whereas in canonical equilibrium turbulence the Reynolds stresses are composed entirely of stochastic, broadband fluctuations, the onset of non-equilibrium organization introduces a distinct, coherent-structural contribution to the total macroscopic stress \cite{Girimaji2024_NJP}. Consequently, rather than conforming to a universal or pre-specified relationship, the co-evolution of the macroscopic structures and the Reynolds stresses is fundamentally dictated by the specific topology of this coherence.

Because the underlying physical mandate of coherence has remained elusive, the fluid dynamics community has historically relied on the identification of visually persistent vortical motions or streaks. To isolate these entities, the field employs a diverse set of exceptionally powerful data-driven and operator-theoretic techniques: Proper Orthogonal Decomposition (POD) identifies variance-maximizing organization; Dynamic Mode Decomposition (DMD) extracts spatiotemporal spectral content; Lagrangian Coherent Structures (LCS) reveal material transport barriers; and resolvent analysis isolates strongly amplified dynamic responses. While these methods are indispensable for reduced-order modeling and kinematic extraction, they are designed to isolate specific mathematical signatures of coherence rather than uncover its fundamental physical origin. Consequently, in contemporary practice, coherence is frequently inferred retrospectively from the characteristics of the chosen detection framework. 
This work aims to shift the paradigm from topological \emph{manifestation} to foundational \emph{causation}, identifying the higher-level organizational principle that dictates the emergence of these structures.

In this work, we aim to address the physical ontology of macroscopic coherence itself. We seek to develop a foundational, non-equilibrium thermodynamics-inspired theory for large-scale organization that is strictly independent of morphological representation---thereby clarifying the exact physical mandate that underlies the emergence of coherence. Thus, the overarching objective is to identify a higher-level organizational principle that explains the genesis of open-flow coherent structures and can serve as the cornerstone of a broadly applicable non-equilibrium theory of turbulence.

The conceptual impetus for this work traces to the seminal investigations of structures in near- and far-from-equilibrium thermodynamic systems, pioneered by Kubo \cite{Kubo1966} and Prigogine \cite{Prigogine1967}. Building upon this legacy, we establish a conceptual bridge that unifies structural emergence in general open thermodynamic systems with those found in turbulent flows. To operationalize this connection, we develop the requisite mathematical machinery by formulating the renormalized Coarse-Grained Navier--Stokes (CGNS) field as a Non-Equilibrium Steady State (NESS). This formulation enables us to systematically classify coherent features based on their degree of departure from equilibrium, ultimately synthesizing a \emph{universal} thermodynamic organizational blueprint for turbulent coherence. By isolating this causal link, we aim to establish a foundational framework for non-equilibrium turbulence theory analogous to the formalism Prigogine achieved for Non-Equilibrium Thermodynamics \cite{Glansdorff1971}. In accordance with this formalism \cite{Glansdorff1971}, the term \emph{thermodynamic} is used throughout this work to describe any physical feature that emerges to accommodate an irreversible throughput imbalance.

Beyond its theoretical significance, this framework possesses profound practical utility. It directly addresses a critical operational challenge in turbulence modeling: determining precisely which scales of motion must be explicitly resolved and which can be reliably parameterized with closure modeling. From this perspective, only those macroscopic structures that mediate large-scale transport beyond the descriptive capacity of canonical equilibrium models mandate direct resolution. Consequently, this study provides a rigorous, physics-based criterion for the optimal computation of complex, non-equilibrium turbulent flows.

To achieve these foundational objectives, we systematically deconstruct the causal, mechanistic, and realized states of macroscopic turbulent organization. Rather than treating coherent structures as statistical accidents, we demonstrate that they arise from a fundamental thermodynamic reorganization required to process the energy throughput mandated by external forcing. We map this thermodynamic necessity onto specific hydrodynamic opportunities, showing that structural emergence is strictly governed by the underlying balance considerations. Depending on the nature of the applied forcing, this emergence manifests either through linear disturbance amplification or macroscopic modal bifurcation. By characterizing the topology-agnostic physical features of this coherent emergence, we ultimately demonstrate how this underlying thermodynamic mandate can be directly employed to guide the optimal, physically rigorous computation of complex turbulent flows.

The theoretical framework developed herein serves as the open-system macroscopic counterpart to the foundational statistical-mechanical formulation of closed, mesoscale (inertial-scale) turbulent systems pioneered by Onsager \cite{onsager1949}, advanced by Kraichnan \cite{Kraichnan1959}, and formalized by Eyink and Sreenivasan \cite{EyinkSreenivasan2006}. Their seminal contributions demonstrated that to satisfy the dissipative anomaly, the inertial cascade must spontaneously generate small-scale, non-differentiable singular structures that act as the intrinsic thermodynamic sink for energy throughput. The framework introduced here directly complements this classical paradigm.

\section{On the Thermodynamic Ontology of Turbulent Structures}

Organized structures are a central feature of many nonequilibrium systems across statistical physics. In thermodynamic systems, these structures emerge to facilitate the processing and dissipation of energy injected by external forcing. The emergence of these structures has been extensively studied in near-equilibrium \cite{KuboTodaHashitsume1991} and far-from-equilibrium systems \cite{Prigogine1977Nobel}. 
This work centers on turbulent flows, which manifest coherent structures---such as streaks, vortices, and rolls---across a diverse range of physical regimes.
As mentioned in the Introduction, the goal of this work is to unify the analytical treatment of structures in both domains, emphasizing the fundamental physical mandate that governs their emergence and ontology.

\subsection{Historical Objections}
\label{sec:historical_objection}

The proposal to interpret turbulent coherent structures as dissipative structures in the sense of Prigogine has historically encountered conceptual resistance within the fluid mechanics and statistical physics communities. It is imperative to clearly enunciate these historical objections, and ensure the current framework successfully addresses them.

These historical objections can be synthesized into three primary categories:
\begin{enumerate}
    \item \textbf{Structural persistence:} Prigogine's classical examples, such as Rayleigh--B\'enard convection cells or chemical clocks, are characterized by spatial rigidity, broken symmetry, and long-term temporal persistence. Conversely, while certain turbulent flows sustain highly persistent macro-structures, many fully developed turbulent fields are dominated by ephemeral features---transient eddies and vortices---that continuously stretch, distort, and break apart. As argued by Anderson and Stein \cite{anderson1984}, many far-from-equilibrium theories mistake highly unstable or chaotic behaviors for true structural entities. From this perspective, a transient turbulent eddy lacks the temporal persistence conventionally required to qualify as a formal dissipative structure.
    \item \textbf{High-dimensionality and the Arrow of Causality:} Prigogine's core mechanism relies on ``order through fluctuations,'' an upward causality where microscopic thermal noise amplifies into macroscopic organization. Fluid dynamicists have historically argued that turbulence operates in the exact opposite direction. According to the classical statistical framework \cite{Kolmogorov1941}, kinetic energy is systematically transferred downward from large macro-flows into increasingly chaotic, smaller-scale fluctuations until it is homogeneously dissipated at the viscous limit. Bricmont \cite{bricmont1995} reinforced this divergence, arguing that Prigogine's terminology does not lead to a constructive mathematical framework capable of resolving high-dimensional fluid chaos. Crucially, this standard objection assumes that all scales of fluid motion must be viewed through a single, continuous top-down cascade.
    
    \item \textbf{Mathematical machinery and order-parameters:} On a strictly mathematical level, Prigogine sought to utilize ``excess entropy production'' as a generalized Lyapunov function to prove the stability of far-from-equilibrium steady states. Turbulence research, however, has so far not successfully identified a universal thermodynamic potential that dictates the evolution of non-equilibrium systems. Without valid mathematical machinery, Prigogine's original thermodynamic approach lacks the quantitative metrics or order parameters required to objectively predict or explain the emergence and stability of turbulent regimes.
\end{enumerate}

\subsection{Ontology of Thermodynamic Structures}

We propose that the structural persistence objection can be resolved by classifying both thermodynamic and turbulent organization based on the nature of the external forcing and the available bifurcation pathways. The specific nature of this organization---whether transient or permanent---depends fundamentally on the system's proximity to equilibrium and its internal dissipative requirements.

\subsubsection*{Near-Equilibrium Thermodynamic Organization}
In the near-equilibrium regime, macroscopic transport under weak, semi-persistent forcing is governed by Kubo linear response theory \cite{KuboTodaHashitsume1991}. While the resulting fluxes are traditionally viewed as smooth averages, the underlying microscopic dynamics are dominated by intermittent, strongly correlated fluctuation events. Within the framework of the Fluctuation--Dissipation Theorem (FDT) \cite{KuboTodaHashitsume1991}, external forcing biases these intrinsic fluctuations---selecting the most probable relaxation trajectories \cite{OnsagerMachlup1953}---to resolve the incipient \emph{thermodynamic crisis} triggered by localized energy throughput imbalances. Crucially, these transient structures do not constitute a new structural branch; their role is strictly \emph{accommodative}, representing a statistical tilt of the equilibrium state rather than a bifurcation. They serve as the physical carriers of the linear response, regressing to the equilibrium state via Onsager's regression hypothesis \cite{Onsager1931_I,Bertini2015} once the local imbalance is relieved. Consequently, while the lifespan of these ephemeral structures is dictated by local fluid dynamics, their emergence is not a mere kinematic artifact; it is driven by a strict \emph{dissipative mandate} to process the injected energy.

\paragraph*{Thermodynamic Exemplars of Transient Structures.}
The primary paradigm is a horizontal fluid layer subjected to a weak, sub-critical adverse temperature gradient ($\Delta T < \Delta T_{\mathrm{crit}}$), existing just beneath the classical Rayleigh--B\'enard convection threshold. In this near-equilibrium regime, the underlying molecular background acts as a chaotic thermal bath. According to the Fluctuation--Dissipation Theorem (FDT) and Onsager's regression hypothesis, spontaneous microscopic fluctuations continuously perturb the system. When a random thermal fluctuation causes a localized packet of fluid to warm, buoyancy forces amplify this packet, temporarily breaking the spatial symmetry of the conduction state to yield an ephemeral, highly correlated macroscopic plume or draft. In due course, the viscous dissipation and thermal diffusivity of this low-Rayleigh flow overwhelm the buoyant amplification---a behavior mathematically mapped via fluctuating hydrodynamics \cite{Zaitsev1971, Lekkerkerker1974, Graham1974} and confirmed experimentally by Berg\'e and Dubois \cite{Berge1976}. The localized throughput surge caused by the temperature fluctuation is successfully processed; the correlated plume enhances heat transport, relaxes the local gradient, and subsequently dissolves back into the chaotic molecular bath. The lifespan of this structure is strictly dictated by the system's linear response function. It does not possess long-term temporal persistence, yet it is undeniably an organized, correlated state. Its transience is not a hallmark of unorganized chaos, but rather the explicit signature of thermodynamic regression returning the system to its stable Non-Equilibrium Steady State (NESS).

Identical archetypes of transient macroscopic organization manifest across diverse physical domains, exemplified by sub-critical Taylor--Couette flows ($Re < Re_{\mathrm{crit}}$) where local gradients filter thermal fluctuations into transient toroidal drafts prior to global instability \cite{Zaitsev1970tc, Snyder1970}. Parallel sub-critical behaviors drive transient organization in autocatalytic chemical reaction--diffusion networks \cite{Vidal1982} and in the orientational alignment of nematic liquid crystals subjected to low voltages \cite{Kai1978}. In all such instances, the system relies on transient, noise-driven macroscopic structures whose explicit thermodynamic mandate is to optimize local transport, relax the imposing gradient, and systematically regress back into the chaotic bath via generalized Onsager relaxation.

\subsubsection*{Far-from-Equilibrium Systems and Dissipative Structures}
Conversely, when a system is driven far from equilibrium by large external gradients, local linear fluxes fail to accommodate the imposed energy throughput. This triggers a global \emph{thermodynamic crisis}, where the rate of energy injection overwhelms the background dissipative capacity. In such states, the structureless \emph{thermodynamic branch} \cite{Prigogine1967} loses global stability \cite{Cross1993,Nicolis1977}, mandating a qualitative topological shift: the system undergoes a symmetry-breaking bifurcation to a highly ordered, macroscopic branch. Unlike transient fluctuations, this persistent organization---formally a \emph{dissipative structure} \cite{Prigogine1967}---represents a fundamental reorganization of the flow. These emergent structures are maintained by continuous forcing and operate at a significantly higher dissipative capacity than the baseline, fulfilling the \emph{dissipative mandate} required to maintain a stable non-equilibrium steady state (NESS).

\paragraph*{Thermodynamic Exemplars of Transformative Structures.}
The classic archetype is a horizontal fluid layer driven beyond the critical threshold by a severe adverse temperature gradient $(\Delta T > \Delta T_{\mathrm{crit}})$. In this far-from-equilibrium regime, the generalized thermodynamic branch (structureless conduction) loses stability entirely \cite{Glansdorff1971}. The system undergoes a macroscopic, symmetry-breaking bifurcation to establish fully realized, stable Rayleigh--B\'enard convection cells. Here, the long-term temporal persistence of the macroscopic rolls is maintained continuously by the strong external forcing. 
As originally conceptualized by Prigogine, these rolls manifest as macroscopic \emph{dissipative structures} driven by thermodynamic necessity; they emerge to accommodate an irreversible energy throughput that strictly exceeds the transport capacity of the baseline structureless state \cite{Prigogine1977Nobel}.
Identical transformative archetypes manifest across parallel disciplines: super-critical Taylor--Couette flows $(Re > Re_{\mathrm{crit}})$ where stable, macroscopically persistent toroidal vortices alter momentum transport; chemical systems operating beyond thermodynamic stability limits via sustained concentration oscillations in Belousov--Zhabotinsky reactions \cite{Nicolis1977}; and electrodynamics via persistent macroscopic convective patterns in liquid crystals subjected to high voltages. In all such instances, the permanent macroscopic reorganization serves as a highly efficient, non-linear pathway to alleviate the intense thermodynamic tension imposed by the environment \cite{Glansdorff1971, Nicolis1977}.

Despite their differing topologies, both near- and far-from-equilibrium structures share a common ontology: they are dynamic organizations that emerge to resolve thermodynamic crises precipitated by energy throughput imbalances.

\begin{table}[t]
\caption{\label{Table1} \textbf{Categorization of organized responses in externally forced thermodynamic systems}. Both regimes emerge to resolve energy throughput imbalances through distinct physical mechanisms.}
\centering
\begin{tabular}{lll}
\toprule
Feature & \shortstack[l]{Near-\\equilibrium} & \shortstack[l]{Far-from-\\equilibrium} \\
\midrule
Framework & Kubo response \& FDT & Prigogine theory \\
Forcing & Weak, semi-persistent & Strong, persistent \\
Manifestation & Transient structures & Reorganization \\
Mechanism & Ephemeral amplification & Symmetry-breaking \\
Role & \textbf{Accommodative} & \textbf{Transformative} \\
\bottomrule
\end{tabular}
\end{table}

\subsection{Bridge to Coherent Structures in Turbulence}

Turbulent flows are open dissipative systems driven by macroscopic forcing (shear, pressure gradients, or disturbances). We hypothesize that turbulent coherent structures are the hydrodynamic realizations of the thermodynamic exemplars established above. By mapping turbulence to this dual-regime framework, we categorize coherent structures into two distinct classes:

\paragraph*{Class I: Accommodative Transient Structures.} 
In many free and wall-bounded shear flows, turbulence remains close to a state of statistical equilibrium where production is, on average, balanced by dissipation and transport. Localized excursions away from this equilibrium give rise to \emph{accommodative} structures dynamically required to nullify the throughput imbalance. While Anderson and Stein \cite{anderson1984} correctly identify such features as ephemeral, we argue they are not chaotic instabilities. Rather, they are explicitly mandated by thermodynamic necessity, serving as the exact hydrodynamic counterparts to the pre-convective fluctuations in Rayleigh--B\'enard systems.

Crucially, the underlying physical system does not undergo a discrete macroscopic bifurcation. Instead, these Class I features operate as continuous, regressive adjustments of the flow field, allowing the system to process statistical throughput variations without transitioning to a new thermodynamic branch. Because the flow lacks an accessible, alternative non-linear solution manifold, it remains firmly anchored to the baseline state. This restriction limits the thermodynamic impact of the external driving, ensuring it remains localized. Thus, while their visual morphology depends on the specific forcing, their thermodynamic function is purely relaxational: they act as the direct open-system manifestations of fluctuations governed by the Fluctuation--Dissipation Theorem (FDT), algebraically dissipating excess variance and smoothly reverting to the mean. Recognizing these structures as FDT extensions directly resolves the historical objection regarding their lack of persistence, as detailed in Section \ref{sec:class1}.

\paragraph*{Class II: Transformative Global Structures.} 
In sharp contrast to near-equilibrium configurations, Class II regimes are fundamentally distinguished by the system's topological access to discrete macroscopic bifurcations. Here, the presence of an accessible bifurcation pathway allows the localized energy throughput to escalate, structurally liberating the external driving to exert a strong, transformative impact on the flow. Under these conditions, the accumulation of localized dissipative tension precipitates a severe structural \emph{thermodynamic crisis} that cannot be resolved by the passive, accommodative mechanisms of Class I structures alone. Instead, the bifurcation pathway permits the system to break away from its baseline statistical equilibrium branch and access alternative, stable non-linear solution manifolds. This process spawns \emph{transformative} dissipative structures---autonomous macroscopic modes that fundamentally reorganize the flow topology and transport pathways to accelerate energy throughput and relieve the accumulated tension. Functioning as the direct hydrodynamic analogues of Prigogine's classical dissipative structures \cite{Prigogine1967}, Class II features do not merely accommodate variations within the existing framework; they self-organize the system into a new, macroscopically distinct branch of the flow physics. An in-depth analysis of these structures is provided in Section \ref{sec:class2}.


\section{Coarse-Graining and the Mathematical Machinery of Coherence}
\label{sec:machinery_coherence}

Having resolved the structural persistence objection through a dual-regime classification, we now address concerns regarding the arrow of causality and the absence of a cohesive mathematical framework. To overcome these, we adopt a rigorous three-layered mathematical presentation.

First, we derive the exact coarse-grained Navier--Stokes (CGNS) equations via spatial filtering, without introducing any modeling assumptions. Second, we introduce scale-dependent closure modeling through approximations of the unresolved stress tensor. This step is representation-dependent and not part of the exact CGNS identity, analogous to simplifying the collision operator in Boltzmann kinetic theory. Third, we define thermodynamic quantities---such as the production-to-dissipation ratio---strictly as diagnostic measures to characterize the nonequilibrium state of the unresolved field. These metrics do not modify the governing equations; rather, they provide a diagnostic signature of the thermodynamic imbalance associated with the emergence of the resolved coherent structures.

The remainder of this section develops these three levels in detail.

\subsection{Coarse-Graining and Emergent Irreversibility}
\label{sec:cgns_ness}

The historical objections regarding dimensionality and irreversibility overlook the foundational methodology of statistical mechanics. Just as Prigogine structures and classical thermodynamic laws emerge from the coarse-graining of hyper-chaotic microscopic molecular collisions via the Boltzmann equation, a similar layered framework must be applied to turbulence coherent structures. 
In classical dissipative structures, the underlying microscopic molecular dynamics are Hamiltonian and time-reversible; irreversibility in Prigogine’s framework emerges only after adopting a mesoscopic, statistically projected description, such as the Boltzmann equation with the \textit{Stosszahlansatz} \cite{Prigogine1967}. Consequently, the arrow of time is not embedded in the microscopic equations themselves, but in the coarse-grained description. 

\begin{figure}[t]
\centering
\begin{tikzpicture}[
  font=\footnotesize,
  >=Stealth,
  box/.style={
    rectangle,
    rounded corners=2pt,
    draw=black,
    line width=0.55pt,
    align=center,
    inner sep=5pt,
    text width=0.78\linewidth
  },
  arr/.style={
    -{Stealth[length=2.5mm,width=1.8mm]},
    line width=0.55pt
  },
  lab/.style={
    font=\footnotesize\itshape,
    fill=white,
    inner sep=1pt
  }
]

\node[box] (forcing) {
\textbf{External forcing / energy injection}\\
Mean shear, pressure gradients, boundary work, body forcing
};

\node[box, below=5mm of forcing] (meso) {
\textbf{Resolved mesoscale dynamics (CGNS / coherent-structure manifold)}\\
Filtered field $U_i$, filtered pressure $\bar{p}$, residual stress
$\tau_{ij}=\overline{u_i u_j}-U_iU_j$
};

\node[box, below=5mm of meso] (turb) {
\textbf{Unresolved turbulent bath (sub-filter central moments)}\\
Incoherent fluctuations $u'_i$ represented by $\tau_{ij}$
};

\node[box, below=5mm of turb] (thermo) {
\textbf{Molecular thermodynamic bath (viscous dissipation / entropy production)}\\
Irreversible conversion of kinetic energy to heat
};

\draw[arr] (forcing) -- (meso)
  node[midway, right=2mm, lab] {$\mathcal{I}_r$};

\draw[arr] (meso) -- (turb)
  node[midway, right=2mm, lab] {$P_u$};

\draw[arr] (turb) -- (thermo)
  node[midway, right=2mm, lab] {$\varepsilon_u$};

\end{tikzpicture}

\caption{\textbf{Hierarchical energy-transfer structure in the coarse-grained Navier--Stokes (CGNS) framework}. External forcing injects power $\mathcal{I}_r$ into the resolved mesoscale dynamics (coherent-structure manifold). A persistent net transfer $P_u$ carries kinetic energy from the resolved manifold to the unresolved turbulent bath represented by the central moments $\tau_{ij}$. The turbulent bath subsequently transfers energy to the molecular thermodynamic bath at rate $\varepsilon_u$, corresponding to irreversible viscous dissipation and entropy production. This schematic represents the renormalization pathway 
$\mathrm{NS} \rightarrow \mathrm{CGNS} \rightarrow \mathrm{RANS}$ and provides a clean entry point for scale-resolving closures (e.g., PANS) through explicit modeling of $P_u$ and $\varepsilon_u$.}
\label{fig:cgns_bath_schematic}
\end{figure}
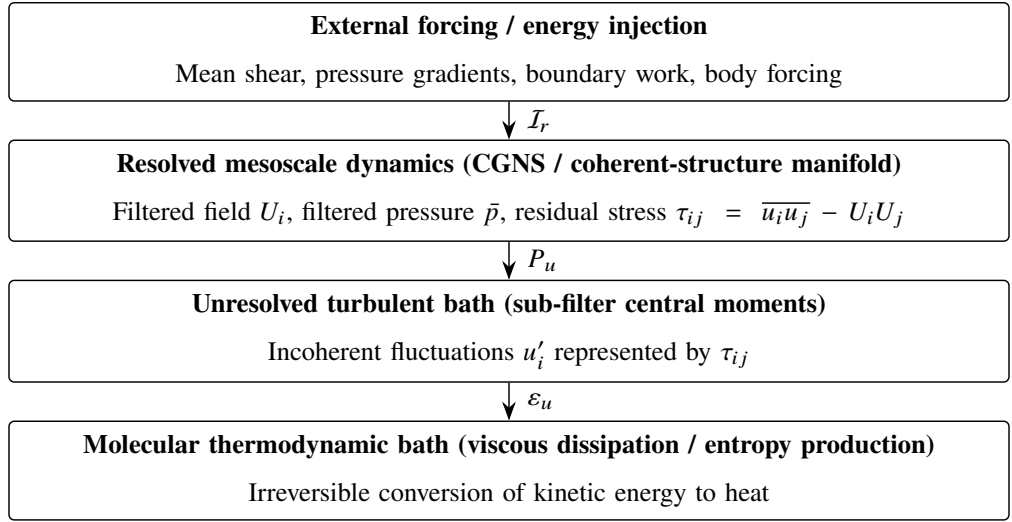

This interpretation of irreversibility carries over to the turbulence scenario quite precisely. At the fully resolved continuum level, the underlying inviscid dynamics (the Euler limit) are deterministic and time-reversible; macroscopic irreversibility emerges  through statistical description and scale demarcation, with the forward cascade providing the effective arrow of time \cite{Frisch1995, Girimaji2024}. This interpretation is consistent with renormalization-group perspectives in turbulence, where unresolved scales are systematically absorbed into effective transport coefficients in coarse-grained descriptions.
For coherent-structure analysis in the spirit of Prigogine, turbulence must therefore be examined through a CGNS description, viewing the flow as a multi-layered hierarchy (Fig. \ref{fig:cgns_bath_schematic}).

The coarse graining can be performed using the Navier--Stokes equations directly or derived from the kinetic Boltzmann equation \cite{girimaji2007}. While the latter approach more clearly illuminates the thermodynamic-hydrodynamic analogy, the former provides the standard entry point for turbulence theory. The hydrodynamic flow field is decomposed into a resolved component $U_i$ and a residual part $u'_i$, with the coherent structures resident in the resolved manifold:
\begin{equation}
\label{decomposition}
    u_i = U_i + u'_i; \quad U_i = \overline{u}_i,
\end{equation}
where the overbar denotes a scale-selective filter. The resulting CGNS equations for $U_i$ are given by \cite{Germano1992}:
\begin{align}
\frac{\partial U_i}{\partial x_i} &= 0, \\
\frac{\partial U_i}{\partial t} + U_j \frac{\partial U_i}{\partial x_j} 
&= -\frac{1}{\rho}\frac{\partial \bar{p}}{\partial x_i} 
+ \nu \nabla^2 U_i 
- \frac{\partial \tau_{ij}}{\partial x_j},
\label{eq:cgns}
\end{align}
where $\bar{p}$ is the filtered pressure and $\tau_{ij}$ is the residual subgrid stress tensor. Critically, this stress is rigorously identified as the generalized \emph{central moment} of the velocity field \cite{Germano1992,Girimaji2006}:
\begin{equation}
\label{sgs_stress}
    \tau_{ij} = \overline{u_i u_j} - U_i U_j.
\end{equation}
This central moment quantifies the irreducible variance of the unresolved degrees of freedom, mediating the inter-scale energy transfer from the resolved manifold to the unresolved turbulent bath.
To close the system, the subgrid stress is modeled via the Boussinesq hypothesis:
\begin{equation}
\label{nu_t} 
    \tau_{ij} - \frac{1}{3}\tau_{kk}\delta_{ij} = -2\nu_u \overline{S}_{ij},
\end{equation}
where $\overline{S}_{ij}$ is the resolved strain-rate tensor and $\nu_u(\mathbf{x})$ is the turbulent eddy viscosity. Much like Onsager's linear force-flux relations \cite{Onsager1931_I}, this closure assumes a linear response between the macroscopic ``force'' ($\overline{S}_{ij}$) and the resulting ``flux'' ($\tau_{ij}$), a condition physically justified for turbulence under local equilibrium \cite{Girimaji2024_NJP}. The key distinction between the Onsager and Boussinesq relations is that the turbulent eddy viscosity is an emergent property of the flow state rather than an intrinsic property of the fluid.

Substituting this closure into Eq. \ref{eq:cgns} and assuming local homogeneity of the filter, the resolved dynamics become:
\begin{equation}
\frac{\partial U_i}{\partial t} + U_j \frac{\partial U_i}{\partial x_j} 
= -\frac{1}{\rho}\frac{\partial \bar{p}}{\partial x_i} 
+ (\nu + \nu_u) \nabla^2 U_i.
\label{eq:cgns2}
\end{equation}
This formulation fundamentally renormalizes the unresolved nonlinear cascade into a macroscopic diffusive mechanism. Consequently, the dissipative environment of the resolved motion is characterized by an effective turbulent viscosity, $\nu_{\text{eff}}(\mathbf{x}) = \nu + \nu_u(\mathbf{x})$. 

It is important to emphasize that the CGNS framework is not a single model, but a resolution-continuous operator hierarchy. In the fully averaged limit, CGNS reduces to the classical Reynolds-Averaged Navier--Stokes (RANS) equations, which define the bath-like macroscopic field. At intermediate resolutions---Scale-Resolving Simulations (SRS) including Large-Eddy Simulations (LES)---CGNS permits the partial liberation of coherent scales from the unresolved bath. In the fully resolved limit, CGNS recovers Direct Numerical Simulations (DNS). Thus, RANS, SRS, and DNS are not distinct paradigms, but limiting realizations of a single renormalized operator.

The utility of the four-layer CGNS hierarchy (Fig. \ref{fig:cgns_bath_schematic}) varies according to the type of coherent structure to be analyzed. For Class I (accommodative) structures, the physics is near-equilibrium. Since these structures are statistically contained within the fluctuations of the bath, a RANS-level description may suffice to capture their mean effects, though increasing resolution enables a more vivid depiction of their accommodative role. This distinction will be further explored in Section IV.

For the Prigogine-type analysis of Class II (transformative) coherent structures, coarse-graining is mandatory even for capturing the most basic statistics of the flow. Because transformative structures represent a symmetry-breaking bifurcation from the turbulent bath, they must be explicitly resolved to be observed. The CGNS framework (Eq. \ref{eq:cgns2}) overcomes the historical objections of dimensionality and irreversibility, serving as a rigorous NESS framework characterized by the following attributes:

\begin{enumerate}
    \item \textbf{Openness:} The system is maintained by sustained external forcing that injects power $\mathcal{I}_r$ at large scales or boundaries.
    \item \textbf{Turbulent bath:} The unresolved scales, parameterized by the central moment $\tau_{ij}$, absorb the cascade. The transfer rate $P_u = -\tau_{ij}\overline{S}_{ij}$ serves as the coarse-grained proxy for irreversible entropy production at the mesoscale.
    \item \textbf{Renormalized chaos:} The filtering operation renormalizes small-scale chaotic dynamics into effective transport terms, enabling a deterministic mesoscale description that retains its far-from-equilibrium character.
    \item \textbf{Stationary reference branch:} While the instantaneous flow is chaotic, the coarse-grained field defines a stationary statistical reference branch suitable for bifurcation and instability analysis.
\end{enumerate}

\subsection{Closure Modeling and the Mathematical Machinery of Coherence}

The emergence and precision of coherent structures are determined by the stability characteristics of the resolved flow (Eq. \ref{eq:cgns2}) and the energetic capacity of the unresolved turbulent bath. As established, the inter-scale energy transfer is modeled via the Boussinesq flux-force relationship (Eq. \ref{nu_t}). The precise empirical form of this closure is secondary to its structural consistency with the underlying subgrid framework. In non-equilibrium pattern formation, emergent macroscopic structures are remarkably insensitive to microscopic detail, provided the governing mesoscale operator respects fundamental symmetries, conservation laws, and irreversible transfer mechanisms \cite{Cross1993,Goldenfeld1992}. Similarly, in kinetic theory, simplified collision operators recover continuum hydrodynamics by enforcing structural irreversibility rather than reproducing exact microscopic collision dynamics \cite{Cercignani1988}. 

With this foundation established, we examine the specific physical constraints required to evaluate the turbulent viscosity, $\nu_u(\mathbf{x})$, such that it faithfully encodes the local non-equilibrium state of the unresolved bath. An admissible closure framework must satisfy the following structural criteria:
\begin{enumerate}
   \item \textbf{Energetic Consistency and Computable Irreversibility:} The exchange of kinetic energy between the resolved macroscopic flow and the unresolved bath must be explicitly and consistently represented. In statistically stationary conditions, the framework must permit a net flux of energy from resolved to unresolved scales, establishing the unresolved field as a dissipative reservoir. Crucially, this irreversible transfer must be closed and computable. The closure must provide explicit representations of unresolved production ($P_u$) and unresolved dissipation ($\varepsilon_u$), enabling quantification of the energy throughput. This throughput is the fundamental metric required to assess the \emph{thermodynamic crisis}—the dissipative bottleneck—that necessitates the emergence and organization of coherent structures.
    \item \textbf{Autonomy of the Bath:} The unresolved subsystem must evolve according to its own dynamical equations, wherein $P_u$ and $\varepsilon_u$ are consistent with the underlying turbulence physics. Because the energy-transfer mediator $\nu_u$ requires the specification of characteristic velocity and time scales, at least a two-equation representation is required to preserve the necessary memory effects of the unresolved field within the mesoscale formulation \cite{Girimaji2024_NJP}.
    \item \textbf{Scale Variability of Coarse-Graining:} The partition between resolved and unresolved scales is not universal, as increasing forcing alters the hierarchy of dynamically active structures. For example, in Rayleigh--B\'enard convection, increasing the Rayleigh number produces qualitatively distinct regimes---from steady rolls to chaotic plumes \cite{Ahlers2009}. Likewise, stochastic thermodynamics demonstrates that the division between resolved entropy production and hidden subgrid contributions depends explicitly on the coarse-graining scale \cite{Esposito2012}. The mesoscale formulation must therefore permit variable resolution so that the coarse-graining threshold can adapt to evolving structural scales.
\end{enumerate}

These requirements, derived from the thermodynamic interpretation advanced in this work, are independent of any particular empirical closure. They establish the strict minimum structural conditions necessary for representing the turbulent bath  capable of supporting coherent organization.

\subsubsection{PANS as the Controlled CGNS Operator}
\label{sec:pans_cgns}

Any scale-resolving simulation (SRS) framework providing a complete two-equation description of the unresolved field can theoretically serve as a CGNS operator. Conversely, zero- and one-equation subgrid closures are intrinsically unsuited as they lack the independent dynamical memory required to autonomously govern the length and time scales of the turbulent bath. While algebraic or dynamic models can adjust test-filter scales, they lack the autonomous temporal history (non-Markovian memory effects) needed to capture the structural phase transitions of a reactive thermodynamic bath.
We adopt the Partially-Averaged Navier--Stokes (PANS) framework \cite{Girimaji2006} as our canonical CGNS operator. PANS retains a two-equation description for the unresolved field while permitting a resolution-continuous description of the macroscopic field $U_i$. The degree of coarse-graining is rigorously specified through the ratios of unresolved-to-total turbulence content:
\begin{equation}
    f_k \equiv \frac{K_u}{K}, \qquad f_{\varepsilon} \equiv \frac{\varepsilon_u}{\varepsilon},
    \label{eq:pans_fkfeps}
\end{equation}
where $\{K, \varepsilon\}$ denote total quantities and $\{K_u, \varepsilon_u\}$ represent the bath components. Crucially, PANS preserves the parent RANS operator as an exact limiting case ($f_k = f_\varepsilon = 1$), acting as a variable-fidelity projection of the underlying physics.

In the PANS methodology, the unresolved eddy viscosity is defined through the dynamically evolving subgrid scales:
\begin{equation}
    \nu_u = C_{\mu}\,\frac{K_u^2}{\varepsilon_u}.
    \label{eq:pans_nuu}
\end{equation}
The unresolved turbulent bath is governed by transport equations for the subgrid energy ($K_u$) and dissipation ($\varepsilon_u$):
\begin{align}
    &\frac{\partial K_u}{\partial t} + U_j\frac{\partial K_u}{\partial x_j} = P_u - \varepsilon_u + \mathcal{D}_{K_u}, \label{eq:pans_ku} \\
    &\frac{\partial \varepsilon_u}{\partial t} + U_j\frac{\partial \varepsilon_u}{\partial x_j} = C_{\varepsilon1}\frac{\varepsilon_u}{K_u}P_u - C_{\varepsilon2}^*\frac{\varepsilon_u^2}{K_u} + \mathcal{D}_{\varepsilon_u}, 
\label{eq:pans_epsu}
\end{align}
where $\mathcal{D}_{K_u}$ and $\mathcal{D}_{\varepsilon_u}$ represent the standard gradient-diffusion transport terms for the bath kinetic energy and dissipation, respectively. To ensure the framework satisfies the resolution-continuity requirement, the model coefficient is dynamically rescaled \cite{Girimaji2006}:
\begin{equation}
    C_{\varepsilon2}^* = C_{\varepsilon1} + \frac{f_k}{f_{\varepsilon}}(C_{\varepsilon2} - C_{\varepsilon1}).
\end{equation}
In this formulation, the unresolved production $P_u = 2\nu_u \overline{S}_{ij}\overline{S}_{ij}$ and dissipation $\varepsilon_u$ are computed directly. This ensures that the Prigogine balance mandated by our previous analysis ($P_u - \varepsilon_u$) appears explicitly as the autonomous source-sink term driving the subgrid state.
This mathematical machinery allows us to treat the turbulent bath not as a static background, but as a reactive environment whose capacity to absorb the cascade determines the stability and liberation of coherent structures.

\subsection{The Thermodynamic Order Parameter}
Based on the energy balance of the bath (Eq. \ref{eq:pans_ku}), we identify the production-to-dissipation ratio of the unresolved scales, 
\begin{equation}
    \Pi \equiv \frac{P_u}{\varepsilon_u},
    \label{eq:order_parameter}
\end{equation}
as the fundamental \textit{order parameter} of turbulent organization. This ratio measures the sustained energetic forcing of the unresolved field relative to its local relaxation capacity. This proposal is formally consistent with Prigogine's \emph{excess entropy production} principle \cite{Prigogine1967}: the ratio $\mathcal{P}_u/K_u$ quantifies the rate of variance injection into the unresolved degrees of freedom, playing a role analogous to the $dQ/T$ metric in non-equilibrium thermodynamics. Normalization by the bath relaxation time ($K_u/\varepsilon_u$) renders this measure dimensionless, yielding $\Pi$ as the "overload" parameter. 

It is crucial to emphasize that $\Pi$ does not function as a classical, rigid order parameter typical of closed equilibrium phase transitions. Because turbulence is an open, highly advective system, $\Pi$ embraces a more flexible interpretation as a localized quantitative measure of non-equilibrium order. Due to the omnipresence of spatial transport, the exact baseline value corresponding to statistical equilibrium is not a universal constant, but dynamically adapts to local convective and diffusive conditions. Nevertheless, in stable Class I configurations, $\Pi$ can be robustly expected to hover near unity—modulated continuously by local transport—whereas it escalates significantly in regions primed for Class II structural emergence, signaling an unmitigated buildup of localized dissipative tension.

In this framework, coherence is not a symptom of unorganized chaos, but a rigorous requirement to resolve the \emph{thermodynamic crisis} precipitated by excessive energy throughput to the bath. As summarized in Table \ref{tab:classes}, the emergence of structures is a deterministic response to the state of the unresolved kinetic bath. Class I structures emerge to accommodate transient excursions where $\Pi > 1$; their subsequent dissolution upon restoring parity ($\Pi \approx 1$) represents the successful fulfillment of their thermodynamic mandate. Conversely, Class II structures arise when the driving force is sustained, necessitating a global bifurcation to reorganize the flow into a new transformative state. These behaviors are examined in detail in Sections IV and V.

\begin{table}[h!]
\centering
\caption{\textbf{Classification of Coherent structures.}}
\label{tab:classes}
\begin{tabular}{@{}lll@{}}
\toprule
\textbf{Parameter} & $\Pi \to 1$ (Regressive) & $\Pi \gg 1$ (Bifurcated) \\ \midrule
\textbf{Attribute} & Class I & Class II \\ 
\textbf{Regime} & Near-Equilibrium & Far-from-Equilibrium \\
\textbf{Exemplar} & Boundary layer streaks & Kelvin--Helmholtz rollers \\
\textbf{Forcing} & Weak, transient & Strong, persistent \\
\textbf{Function} & Local accommodation & Global reorganization \\  
\bottomrule
\end{tabular}
\end{table}


\section{Class I Coherence: Near-Equilibrium Accommodative Response}
\label{sec:class1}

In turbulence, infinite-time spectral instability is sufficient but not strictly necessary for structure formation. 
For example, in wall-bounded turbulence, the emergence of macroscopic organization does not necessarily imply a global transition to a new thermodynamic branch. 
Instead, these Class I coherent structures represent \emph{accommodative responses}---cyclically regenerating, non-modal excursions that arise to resolve localized throughput imbalances within a spectrally stable NESS \cite{Trefethen1993, Farrell1993}. To bridge this phenomenon to our thermodynamic framework, we evaluate these structures as the response of a weakly forced system using Kubo’s Fluctuation-Dissipation Theorem (FDT). This section follows a deductive progression to solidify this characterization: we establish the mathematical character of Class I coherence as a generalized susceptibility problem. This paradigm is then illustrated through the physical exemplar of the near-wall autonomous cycle, showing how structures act to restore thermodynamic parity. Finally, we provide a computational treatment using the CGNS PANS framework to demonstrate that the fidelity of captured coherence is fundamentally tied to the explicit resolution of the driving fluctuations.

\subsection{Mathematical Character: Transient Amplification}
\label{sec:class1_fdt}

To capture the response of a near-equilibrium turbulent flow to weak forcing, we decompose the resolved CGNS velocity into a statistically stationary base state and a coherent macroscopic perturbation: $U_i(\mathbf{x},t) = \langle U_i \rangle(\mathbf{x}) + U^{cs}_i(\mathbf{x},t)$. Substituting this into the CGNS evolution equation (Eq. \ref{eq:cgns2}) and linearizing yields a forced system. The mean convective and viscous terms---crucially including the effective turbulent viscosity $\nu_{\text{eff}} = \nu + \nu_u$ derived in Eq. \ref{eq:cgns2}---combine to form the linear Navier-Stokes operator $\mathcal{L}$, which embeds the non-normal amplification potential of the mean profile. 

While the Boussinesq closure successfully captures the mean dissipative action of the unresolved bath---yielding the smooth, effective-viscosity dynamics of Eq.~\ref{eq:cgns2}---the actual unresolved stress $\tau_{ij}$ inherently contains rapid fluctuations around this mean Boussinesq state. These fluctuations act as a continuous forcing field $f_i$ \cite{Farrell1993}:
\begin{equation}
    f_i(\mathbf{x},t) = \frac{\partial (\tau_{ij}^{\text{Bouss}}-\tau_{ij})}{\partial x_j}.
\end{equation}
The evolution of the coherent perturbation $U^{cs}_i$ from this base state subject to the forcing is thus governed by:
\begin{equation}
\label{eq:Kubo_eqn}
    \frac{\partial U^{cs}_i}{\partial t} = (\mathcal{L}U^{cs})_i + f_i.
\end{equation}
Expressing the coherent perturbation and the forcing as Fourier integrals transforms the system into the frequency domain:
\begin{equation}
    i\omega \hat{U}^{cs}_i = (\mathcal{L}\hat{U}^{cs})_i + \hat{f}_i.
\end{equation}
Rearranging this relates the coherent structural response $\hat{U}^{cs}_i$ to the forcing field $\hat{f}_j$ via the resolvent operator:
\begin{equation}
    \hat{U}^{cs}_i(\mathbf{x},\omega) = (\mathcal{H}(\omega)\hat{\mathbf{f}}(\mathbf{x},\omega))_i, \quad \text{where} \quad \mathcal{H}(\omega) = (i\omega \mathbf{I} - \mathcal{L})^{-1}.
\end{equation}

This formulation constitutes a rigorous application of Kubo's fluctuation-dissipation theorem (FDT) \cite{Kubo1966} along the lines of fluctuating hydrodynamics \cite{Zaitsev1971, Lekkerkerker1974, Graham1974}. The notable difference is that the operator $\mathcal{L}$ here represents the linear operator of the renormalized CGNS equation rather than a molecular transport matrix. Consequently, this framework directly operationalizes our Class I thermodynamic ontology: just as sub-critical thermal fluctuations are filtered into transient macroscopic organization in near-equilibrium thermodynamic systems, the rapid unresolved subgrid fluctuations are filtered by $\mathcal{L}$ to generate the accommodative, transient coherent perturbations described herein.

The resolvent operator $\mathcal{H}(\omega)$ acts exactly as the \emph{generalized susceptibility} of the coarse-grained system. Because the underlying linearized operator $\mathcal{L}$ of turbulent shear flows is highly non-normal, this generalized susceptibility is massively skewed. It selectively filters and amplifies the externally driven forcing $f_i$ into organized, macroscopic coherent motion. 
Within this framework, the widely successful \emph{Resolvent Analysis (RA)} approach \cite{McKeonSharma2010} emerges as a specific hydrodynamic implementation of this broader linear response theory. By mapping transient growth onto Kubo's FDT, we demonstrate that Class I coherence is not a sign of spectral instability, but a cyclically regenerating thermodynamic relief valve necessitated by the continuous action of the externally imposed shear forcing.

\begin{table}[t]
\small
\centering
\caption{\textbf{Conceptual and mathematical isomorphism}.  Classical Rayleigh-B\'enard thermodynamic threshold and the present turbulence framework for Class I accommodative structures.}
\label{tab:isomorphism}
\begin{tabular}{lll}
\hline \hline
\parbox[t]{1.3in}{\raggedright \textbf{System}} & 
\parbox[t]{2.6in}{\raggedright \textbf{Near-equilibrium thermodynamic (RB)}} & 
\parbox[t]{2.6in}{\raggedright \textbf{Near-equilibrium turbulence (Class I)}} \\
\hline
\parbox[t]{1.3in}{\raggedright \textbf{Nature of the Bath}} & 
\parbox[t]{2.6in}{\raggedright Microscopic and thermal.} & 
\parbox[t]{2.6in}{\raggedright Hydrodynamic and turbulent.} \\
\hline
\parbox[t]{1.3in}{\raggedright \textbf{Amplification\\ Mechanism}} & 
\parbox[t]{2.6in}{\raggedright Localized adverse density gradient acting on thermal fluctuations.} & 
\parbox[t]{2.6in}{\raggedright Non-normal mean velocity shear driving non-modal transient growth.} \\
\hline
\parbox[t]{1.3in}{\raggedright \textbf{Linear Response\\ Matrix}} & 
\parbox[t]{2.6in}{\raggedright Symmetrized, normal Navier-Stokes operator.} & 
\parbox[t]{2.6in}{\raggedright Highly skewed, non-normal resolvent operator ($\mathcal{H}(\omega)$).} \\
\hline
\parbox[t]{1.3in}{\raggedright \textbf{System Regulation}} & 
\parbox[t]{2.6in}{\raggedright Constant, passive molecular transport coefficients.} & 
\parbox[t]{2.6in}{\raggedright Dynamic subgrid environment regulated by local imbalance ($\Pi_{\text{loc}}$).} \\
\hline
\parbox[t]{1.3in}{\raggedright \textbf{Structural Lifecycle}} & 
\parbox[t]{2.6in}{\raggedright Ephemeral plumes maximizing heat transport to restore conduction state.} & 
\parbox[t]{2.6in}{\raggedright Transient streaks and hairpins maximizing momentum transport to restore shear state.} \\
\hline \hline
\end{tabular}
\end{table}

\subsection{Turbulence Exemplar: Near-Wall Autonomous Cycle}
\label{sec:class1_exemplar}

To illustrate Class I structures in turbulence, we examine the canonical example of wall-bounded turbulence. In this regime, the mean shear serves as the localized amplification mechanism, playing a role directly analogous to the adverse density gradient (buoyancy) in the sub-critical Rayleigh-B\'enard (RB) thermodynamic exemplar.
Much like the sub-critical RB thermal field, the log-law region typically maintains a statistically stationary state where turbulent production and dissipation are balanced on average ($\Pi \approx 1$). Because this highly sheared turbulent mean state lacks a global symmetry-breaking bifurcation, the emergence of macroscopic structures does not represent a transition to a new persistent thermodynamic branch. Rather, Class I structures emerge as the exact dynamically favored \emph{accommodative responses} ($\hat{U}^{cs}_i$) predicted by the generalized susceptibility $\mathcal{H}(\omega)$. The conceptual and mathematical isomorphism between these near-equilibrium thermodynamic and turbulent systems is detailed in Table~\ref{tab:isomorphism}. 

Extensive DNS data of channel and boundary-layer flows~\cite{kim1987turbulence, moser1999direct, schlatter2010assessment, lozano2014effect} provide robust evidence for this transient, regressive organization, which maps onto our thermodynamic framework through the following cyclical stages:

\paragraph*{1. Stochastic Gradient Intensification.} 
Weak streamwise vortices, acting as the endogenous noise $f_i$ from the bath, undergo non-modal \emph{lift-up} amplification. This process is governed by the skewed susceptibility of the mean shear, encoded within the resolvent operator $\mathcal{H}$ \cite{Trefethen1993}. The resulting redistribution of mean momentum into elongated low-speed streaks drives a surge in turbulent production, localized specifically beneath the vortex legs \cite{kim1987turbulence, moser1999direct}. Because the corresponding dissipation remains more uniformly distributed, this spatial disparity precipitates a \emph{local throughput imbalance} ($\Pi_{\text{loc}} > 1$).

\paragraph*{2. Localized Structural Response.}
When $\Pi_{\text{loc}}$ exceeds its equilibrium value of unity, the background turbulence becomes locally insufficient to accommodate the elevated production rate. To mitigate this thermodynamic bottleneck, the flow triggers a rapid transient amplification of targeted structural modes as dictated by the operator $\mathcal{H}$ \cite{McKeonSharma2010}. Secondary instabilities act as the catalyst, breaking the intensified streaks into coherent vortical structures, such as hairpins \cite{Hamilton1995, Waleffe1997}.

\paragraph*{3. Non-Equilibrium Flux Redistribution.}
These emergent coherent vortices generate intense \emph{sweeps} and \emph{ejections} \cite{Wallace1972,Willmarth1972}. By rapidly transporting momentum, these coherent motions relax the local mean shear and extinguish the production surge. Thus, transient structures emerge as the dynamically selected optimal responses of the underlying fluctuation field to alleviate the local throughput imbalance. 

\paragraph*{4. Structure Dissolution and Restoration of the Bath.}
As the throughput imbalance is alleviated and $\Pi_{\text{loc}}$ returns to unity, the structures exhaust their thermodynamic mandate. By successfully redistributing momentum, the vortices deplete the very background shear that fueled their growth \cite{Jimenez1991, Waleffe1997}. Deprived of this sustenance, the structures rapidly decay. Following Onsager's regression hypothesis \cite{onsager1949}, these organized fluctuations dissolve back into the background cascade, restoring the flow to an unorganized bath until the next noise-driven cycle.

\subsection{CGNS as a Resolution-Continuous Operator}
\label{sec:class1_computation}

Because Class I coherent structures are fundamentally regressive---arising strictly to alleviate localized throughput imbalances before dissolving back into the cascade---they do not permanently alter the global dynamic branch of the flow. As this is a near-equilibrium flow, even fully averaged closures like RANS ($f_k=1$) can reliably predict the stationary mean field without explicitly resolving the transient coherent structures. This thermodynamic reality explains why resolvent analysis \cite{McKeonSharma2010}, which utilizes this fully averaged description as the baseline flow, is successful in predicting the potential for structural amplification. 

\begin{figure}[t]
    \centering
    \includegraphics[width=0.85\textwidth]{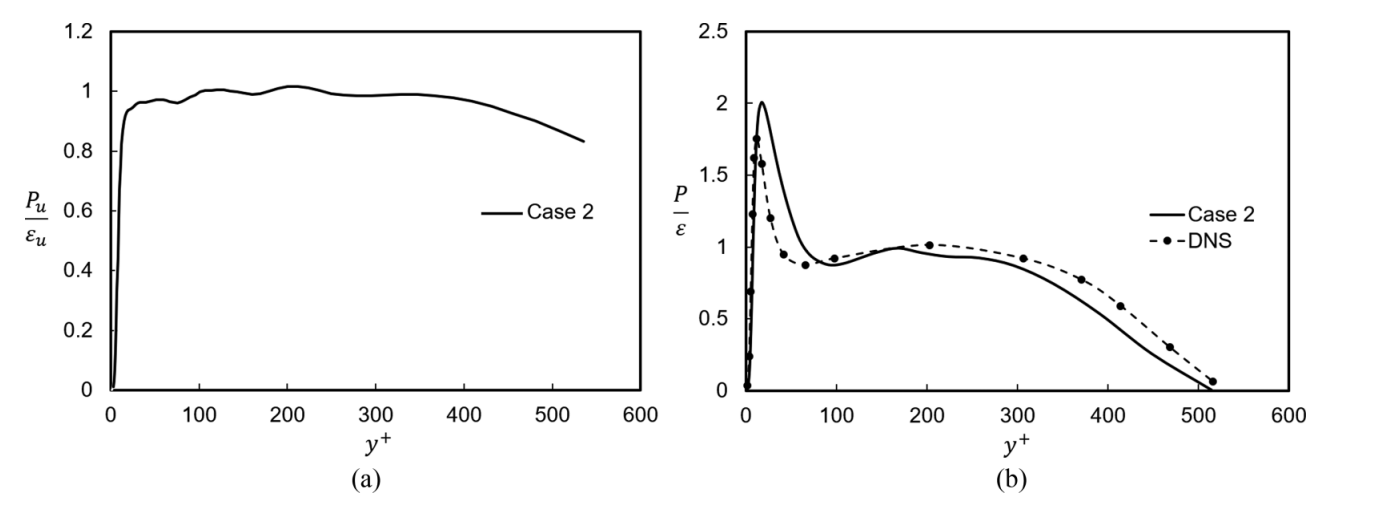}
    \caption{\textbf{
    Production--dissipation balance in equilibrium wall-bounded turbulence}.
    (a) PANS results showing that the unresolved production-to-dissipation ratio,
    $P_u/\varepsilon_u$, remains close to unity across different resolution
    levels in the logarithmic layer of an equilibrium boundary layer.
    (b) Corresponding RANS predictions, compared with experimental data, indicating
    a comparable near-equilibrium balance in the fully averaged limit.
    The persistence of $P_u/\varepsilon_u \approx 1$ across modeled and
    partially resolved descriptions supports the interpretation of both RANS and
    coarse-grained PANS fields as representations of the statistically stationary
    nonequilibrium steady-state (NESS) flow.
    }
    \label{poverepsilon}
\end{figure}

Detailed CGNS computations of turbulent boundary layers using the Partially Averaged Navier-Stokes (PANS) framework at different degrees of resolution have been performed in recent studies \cite{TazraeiGirimaji2019, Kambleetal2022}.
Using standard boundary layer scaling arguments, Tazraei and Girimaji \cite{TazraeiGirimaji2019} demonstrate that the unresolved production-dissipation balance ($P_u = \varepsilon_u$) must be sustained at different levels of coarse graining. A fundamental requirement for any CGNS operator in this near-equilibrium regime is the preservation of the baseline energetic equilibrium.
PANS successfully respects this mandate, capturing and maintaining this near-equilibrium balance across all degrees of physical resolution (Fig.~\ref{poverepsilon}).

Despite this shared ability to maintain the mean thermodynamic balance, RANS and scale-resolving PANS exhibit distinct capabilities for capturing the coherent structures themselves. While RANS adequately predicts the stationary mean field, it inherently cannot capture the time evolution of the structural relief valve. To capture these dynamically evolving transient states, scale-resolving approaches are required. 

Crucially, for these modally stable Class I flows, the PANS framework behaves as a \textbf{resolution-continuous operator} \cite{Kambleetal2022}. The fidelity and precision of the captured transient structures are directly tied to the degree of resolution: as the unresolved kinetic energy fraction ($f_k$) is lowered, the accuracy of the resolved coherent structures progressively increases \cite{TazraeiGirimaji2019}. This occurs because, with decreasing $f_k$, a greater portion of the endogenous noise $f_i$ (identified in Section \ref{sec:class1_fdt}) is explicitly resolved rather than modeled \cite{TazraeiGirimaji2020}. This continuous liberation of the forcing field allows the time evolution of the captured coherent structures to improve correspondingly, providing a progressively sharper lens through which the flow's inherent accommodative structures are resolved.

\section{Class II Structures: Transformative Organization Far-from-Equilibrium}
\label{sec:class2}

Class II coherent structures represent a fundamental transition in the flow's dynamic state, mirroring the paradigm of classical Prigogine theory \cite{Prigogine1967, Nicolis1977}. These transformative structures emerge when a system is driven sufficiently far from equilibrium that the background turbulent bath can no longer satisfy the global throughput requirements via simple non-modal intensification. At this critical juncture, the dissipative capacity of the turbulent bath becomes overwhelmed, triggering a symmetry-breaking bifurcation to a highly organized structural branch capable of accommodating the excess energy throughput. Under the burden of a severe production--dissipation imbalance, the emergent Class II structures radically reconfigure the flow by establishing a distinct dynamical regime that fundamentally redefines macroscopic turbulent transport and dissipation.

We begin by examining the mathematical character of Class II structures, shifting from the non-modal susceptibility of the mean flow to the global modal instabilities that signal a branch transition. This is followed by a physical exemplar---the cylinder wake---which demonstrates how the emergence of the K\'arm\'an vortex street serves as a thermodynamic necessity to alleviate global throughput overload. Finally, we present CGNS results using the PANS framework to demonstrate that, unlike the resolution-continuous nature of Class I flows, Class II structures require a distinct spectral bifurcation to be captured, marking the limit where fully averaged closures fundamentally fail.

\subsection{Mathematical Character: Global Instability}
\label{sec:class2_math}

To analyze the onset of transformative coherence, we once again decompose the flow into the unbifurcated base state and a macroscopic coherent perturbation: $U_i(\mathbf{x},t) = \langle U_i \rangle(\mathbf{x}) + U^{cs}_i(\mathbf{x},t)$. Substituting this decomposition into the CGNS equations and linearizing about the NESS baseline yields the autonomous evolution problem:
\begin{equation}
\frac{\partial U^{cs}_i}{\partial t} = (\mathcal{L}\mathbf{U}^{cs})_i,
\end{equation}
where the effective dissipation of the turbulent bath ($\nu_{\text{eff}}$) is strictly embedded within the linear operator $\mathcal{L}$. This ensures kinematic and thermodynamic consistency: to survive, the coherent perturbation must overcome the dissipative resistance of the highly diffusive, unbifurcated NESS field. 

Crucially, notice the absence of the endogenous forcing function ($f_i$) in this linearized formulation. This mathematical omission highlights the fundamental physical distinction between the two regimes, which is governed entirely by the spectral nature of $\mathcal{L}$. In Class I flows, the system acts as a noise-driven \textit{amplifier}; endogenous forcing is the sole driver of accommodative structures via the operator's non-normal susceptibility, making the continuous inclusion of $f_i$ mandatory. Conversely, Class II structures arise from a global modal instability. In this regime, at least one eigenvalue of $\mathcal{L}$ possesses a positive real part ($\sigma > 0$), rendering the baseline NESS linearly unstable. Once the critical threshold is crossed, the coherent perturbation extracts energy directly from the macroscopic base state. The system transitions into an autonomous \textit{oscillator}, exponentially dominating the dynamics and rendering the background endogenous fluctuations mathematically secondary and structurally inconsequential. This autonomous growth of $U^{cs}_i$ represents a mandatory symmetry-breaking bifurcation, signaling the system's transition to a more efficient organizational branch to satisfy the global throughput mandate.

\subsection{Physical Exemplar: Cylinder Wake Flow}
\label{sec:class2_physics}

To ground the foregoing theoretical construction, we now examine a canonical manifestation of transformative Class II coherence: the wake of a circular cylinder. We analyze this system as the fluid-dynamic, shear-driven analogue of Prigogine's canonical buoyancy-driven exemplar: Rayleigh--B\'enard convection (RBC). By tracing their parallel evolutions, we propose that wake coherent structure organization is not incidental morphology, but an absolute structural response mandated by sustained irreversible throughput in an open system.

Both the cylinder wake and RBC exhibit a hierarchy of layered instabilities as their respective control parameters (Reynolds number, $Re$, and Rayleigh number, $Ra$) increase. Crucially, we must distinguish between primary bifurcations that occur in the pre-turbulent laminar regime and secondary mesoscale structures that emerge strictly to enable a fully developed turbulent bath to resolve a local thermodynamic crisis.

\paragraph*{1. The Pre-Turbulent Dissipative Cascade ($Re \lesssim 260$ / $Ra \lesssim 10^5$).}
At low forcing levels ($Re < 47$ and $Ra < Ra_c$), both systems accommodate energy throughput entirely via microscopic molecular diffusion—viscous momentum diffusion in the wake, and pure thermal conduction in RBC. When macroscopic forcing locally exceeds this diffusive capacity, the systems undergo a primary global bifurcation. The cylinder wake experiences a supercritical Hopf bifurcation ($Re \approx 47$) generating the periodic K\'arm\'an vortex street, perfectly mirrored by the critical pitchfork bifurcation ($Ra \approx 1708$) in RBC that spawns steady 2D convection rolls. As throughput continues to increase, these primary topological pathways saturate. To process the excess energy, both systems exploit nonlinear mode coupling to pump energy into higher spatial and temporal harmonics \cite{Schmid2010}. Ultimately, these two-dimensional states become unstable to 3D perturbations, yielding the spanwise Mode A and Mode B transitions in the cylinder wake \cite{BarkleyHenderson1996,Williamson1996} and 3D cross-roll or oscillatory instabilities in RBC. From a thermodynamic perspective, this entire sequence—from the pure diffusion regime through 3D breakdown—constitutes a cascading chain of pre-turbulent dissipative structures strictly mandated by increasing irreversible throughput.

\paragraph*{2. Establishment of the Turbulent NESS ($Re \sim 300$).}
Following the 3D breakdown, both flows attain a fully turbulent state. By $Re \sim 300$, the cylinder wake's velocity field comprises a stationary mean-flow component, the primary global shedding, and a continuous spectrum of smaller-scale chaotic fluctuations. This parallels the onset of ``soft turbulence'' in RBC. In this regime, kinetic energy extracted from the macroscopic forcing is transferred through coherent motions and the incoherent bath before being irreversibly dissipated. This fluctuation-mediated state constitutes the hydrodynamic NESS field: a statistically stationary configuration in which production balances dissipation ($\mathcal{P} \approx \varepsilon$) and energy throughput is successfully accommodated. Mathematically, the velocity field in this baseline NESS is captured by a triple multiscale decomposition:
\begin{equation}
    \mathbf{U}(\mathbf{x},t) = \langle \mathbf{U} \rangle(\mathbf{x}) + \mathbf{U}^{vs}(\mathbf{x},t) + \mathbf{u}^{tb}(\mathbf{x},t),
\end{equation}
where $\langle \mathbf{U} \rangle$ is the bulk background field, $\mathbf{U}^{vs}$ is the autonomous coherent field due to global vortex shedding, and $\mathbf{u}^{tb}$ represents the structureless turbulent bath.

\paragraph*{3. Mesoscale Thermodynamic Crisis and Higher-Order Emergence ($Re \sim 10^3$).}
As $Re$ approaches the Bloor transition ($\mathcal{O}(10^3)$), the separated shear layers developing directly off the cylinder exhibit increasingly sharp inflectional profiles, with rapidly decreasing shear-layer thickness \cite{Bloor1964,Norberg1987}. At this critical juncture, the established turbulent NESS becomes locally insufficient to process the intense shear gradient through its baseline cascade mechanisms. Faced with this mesoscale thermodynamic crisis, the system mandates a higher-order structural response: the separated shear layer undergoes a Kelvin--Helmholtz (KH) instability, dynamically reorganizing into mesoscale vortical rollers. The velocity field thus expands into a quadruple multiscale decomposition:
\begin{equation}
    \mathbf{U}(\mathbf{x},t) = \langle \mathbf{U} \rangle(\mathbf{x}) + \mathbf{U}^{vs}(\mathbf{x},t) + \mathbf{U}^{KH}(\mathbf{x},t) + \mathbf{u}^{tb}(\mathbf{x},t),
\end{equation}
where $\mathbf{U}^{KH}$ represents the newly mandated, higher-order Class II structures explicitly generated to relieve the localized dissipative tension. Sustained by local mean-shear production via an autonomous global instability ($\sigma > 0$), these KH rollers fundamentally reorganize the irreversible pathways, shifting the spatial distribution of turbulent dissipation $\varepsilon(\mathbf{x},t)$ upstream. Like localized thermal plumes in high-Rayleigh-number convection, the KH rollers represent a secondary coherent response triggered explicitly within an already established turbulent NESS to maintain the cascade under extreme localized forcing.

Overall, Class II turbulent coherent structures are very similar to Prigogine coherent structures as shown in Table \ref{tab:classII_isomorphism}.

\begin{table}[t]
\small
\centering
\caption{\textbf{Conceptual and mathematical isomorphism}. Far-from-equilibrium thermodynamic systems (super-critical Rayleigh-B\'enard) and the present turbulence framework for Class II transformative structures.}
\label{tab:classII_isomorphism}
\begin{tabular}{lll}
\hline \hline
\parbox[t]{1.3in}{\raggedright \textbf{System}} & 
\parbox[t]{2.6in}{\raggedright \textbf{Far-equilibrium thermodynamics}} & 
\parbox[t]{2.6in}{\raggedright \textbf{Far-equilibrium turbulence}} \\
\hline
\parbox[t]{1.3in}{\raggedright \textbf{Nature of the Bath}} & 
\parbox[t]{2.6in}{\raggedright Microscopic and thermal.} & 
\parbox[t]{2.6in}{\raggedright Hydrodynamic and turbulent.} \\
\hline
\parbox[t]{1.3in}{\raggedright \textbf{Amplification\\ Mechanism}} & 
\parbox[t]{2.6in}{\raggedright Severe adverse density gradient driving pitchfork global bifurcation.} & 
\parbox[t]{2.6in}{\raggedright Inflection in macroscopic shear driving a Hopf global bifurcation.} \\
\hline
\parbox[t]{1.3in}{\raggedright \textbf{System Regulation}} & 
\parbox[t]{2.6in}{\raggedright Symmetry-breaking order accelerating transport and ultimately dissipation.} & 
\parbox[t]{2.6in}{\raggedright KH rollers fundamentally redefining macroscopic transport and dissipation.} \\
\hline
\parbox[t]{1.3in}{\raggedright \textbf{Structural Lifecycle}} & 
\parbox[t]{2.6in}{\raggedright Transformative structures represent a new stable state.} & 
\parbox[t]{2.6in}{\raggedright Transformative structures as primary agents extracting excess throughput.} \\
\hline \hline
\end{tabular}
\end{table}

\subsection{CGNS as a Bifurcation Operator}
\label{sec:operationalizing}

Having established the physical manifestations of the dissipation mandate, we now utilize CGNS-PANS results from Pereira \textit{et al.} \cite{Pereira2018JCP} for a cylinder wake at $Re=3,900$ to map these abstract concepts onto observed flow physics. We employ the unresolved kinetic energy fraction, $f_k$, as a strict coarse-graining control parameter and systematically vary the resolution from the RANS limit ($f_k = 1.0$) down to a highly resolved state ($f_k = 0.25$).

\begin{figure}[htbp]
\centering
\includegraphics[width=\columnwidth, trim={0cm 1cm 0cm 1cm}, clip]{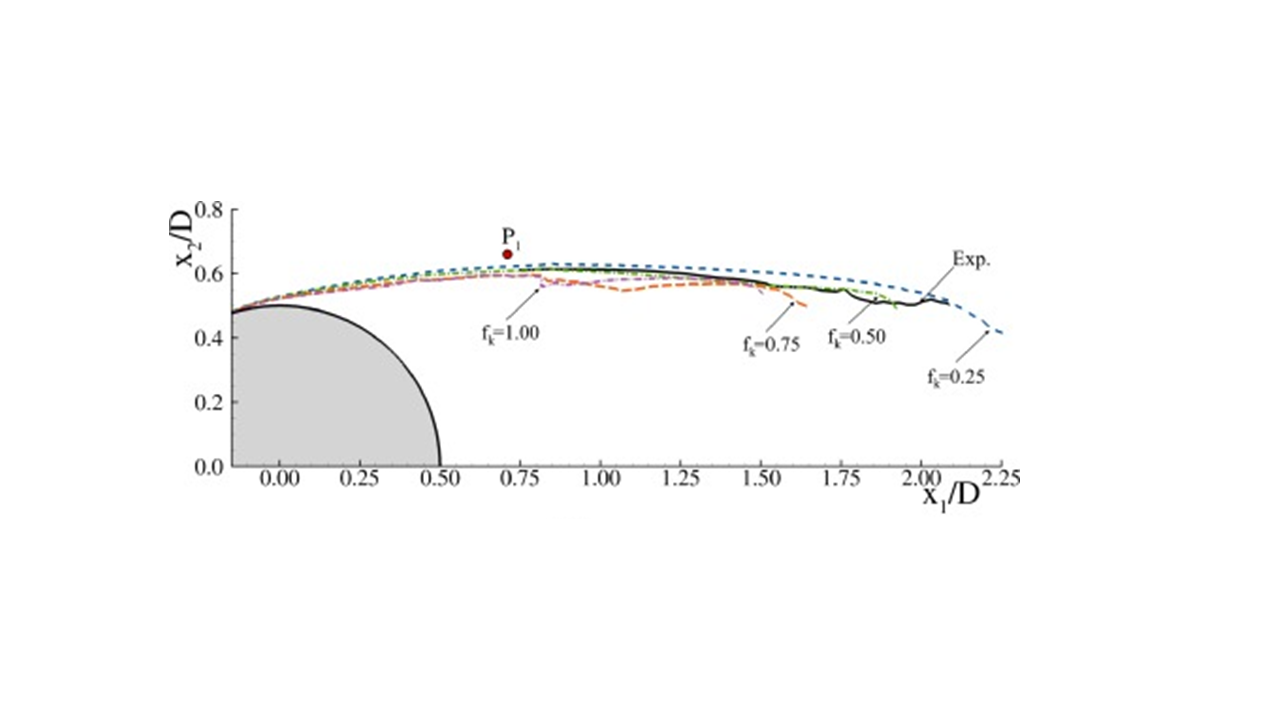}
\vspace{-18pt}
\caption{\textbf{
Mean-flow inflection line in the wake of a circular cylinder at $Re=3,900$}. Results from PANS simulations at varying coarse-graining levels ($f_k = 1.0$ to $0.25$) are compared against experimental data \cite{Pereira2018JCP}. The red marker $P_1$ denotes the characteristic location for the onset of Kelvin-Helmholtz instabilities. The robustness of the inflectional topology across $f_k$ resolutions and closure models supports the use of the RANS field ($f_k=1.0$) as a representative NESS for examining the emergence of coherent structures.
}
\label{fig:inflection_line_RANS}
\end{figure}

In this framework, the baseline NESS represents the stationary flow state existing just prior to the onset of the Kelvin-Helmholtz (KH) instabilities. Figure~\ref{fig:inflection_line_RANS} plots the mean-flow inflection line in the wake, which dictates the spatial origin and early development of the KH structures. The results demonstrate that this inflectional topology is captured with high fidelity across all $f_k$ resolutions when compared to experimental data. Furthermore, as noted in \cite{Pereira2018JCP}, this prediction remains robust across different subgrid closures (e.g., $k\text{--}\omega$ and SST). 

The main utility of the NESS is to predict the \textit{onset} of a structure, not its final saturated state. We propose that a fully averaged RANS flow field—though incapable of predicting the fully developed wake—provides an analytically sound proxy for this unbifurcated NESS. Because the RANS computation accurately captures the mean-flow topology upstream of the structural onset, it reliably identifies the spatial origin of the instabilities (the KH rollers denoted by $P_1$ in Fig.~\ref{fig:inflection_line_RANS}). Thus, it is adequately suited to serve as the baseline for predicting the mandatory emergence of Class II coherent structures. The known spatial inaccuracy of RANS downstream of this structural emergence does not diminish its utility as an onset predictor.

We now turn to characterizing the saturated state in terms of the throughput imbalance quantified by the order parameter $\Pi = P_u/\varepsilon_u$. This parameter is analytically linked to the unresolved timescale ratio \cite{Taghizadehetal2020}:
\begin{equation}
    \frac{P_u}{\varepsilon_u} \approx 0.09 \left( \frac{S K_u}{\varepsilon_u} \right)^2.
\end{equation}
Crucially, the value $S K_u/\varepsilon_u \approx 3.3$ represents the local equilibrium limit ($\Pi \approx 1$). In a numerical simulation, exceeding this threshold creates a severe \textit{dissipative tension}. If the structures are not resolved, the closure model attempts to process the excess production by generating an unphysically high eddy viscosity ($\nu_u$). This artificially lowers the effective computational Reynolds number, $Re_{eff} = UD/(\nu + \nu_u)$, thereby suppressing the physically needed bifurcations whose emergence depends on crossing a critical Reynolds number threshold. With this criterion in mind, we now analyze the results from PANS simulations across four resolution levels.

\begin{figure*}[t]
    \centering
    \subfloat[Dissipative Tension: $S K_u / \varepsilon_u$ contours. \label{fig:tension}]{
        \includegraphics[width=0.48\textwidth]{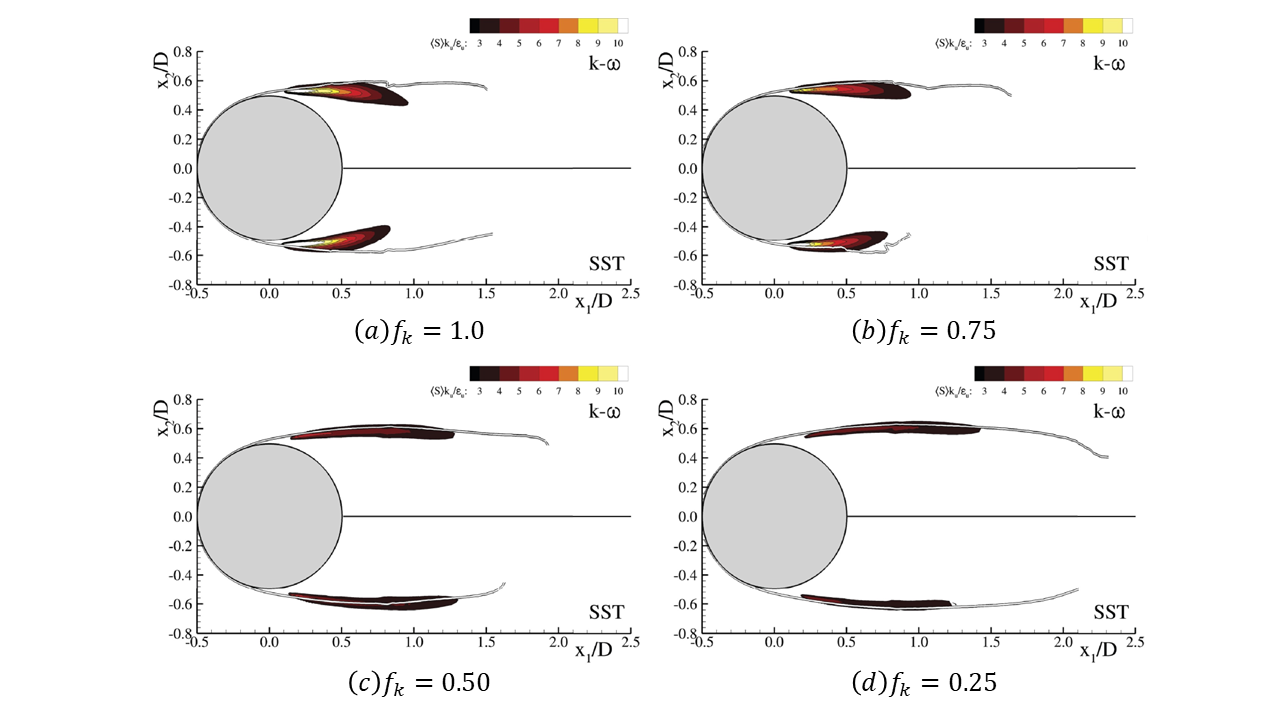}
    }
    \hfill
    \subfloat[Effective Reynolds Number (Bloor Limit check). \label{fig:bloor}]{
        \includegraphics[width=0.48\textwidth]{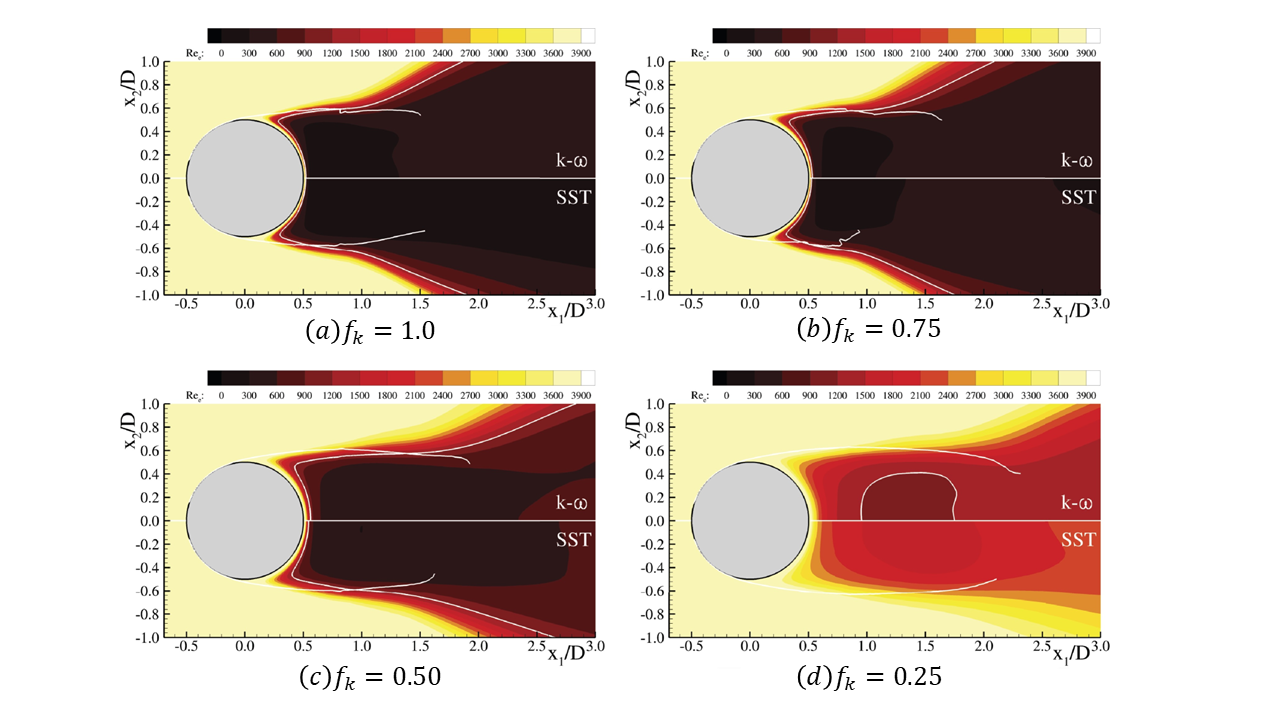}
    } \\ 
    \vspace{10pt} 
    \subfloat[Energy Spectra: $f_{KH}$ vs. VS harmonics. \label{fig:spectra}]{
        \includegraphics[width=0.55\textwidth]{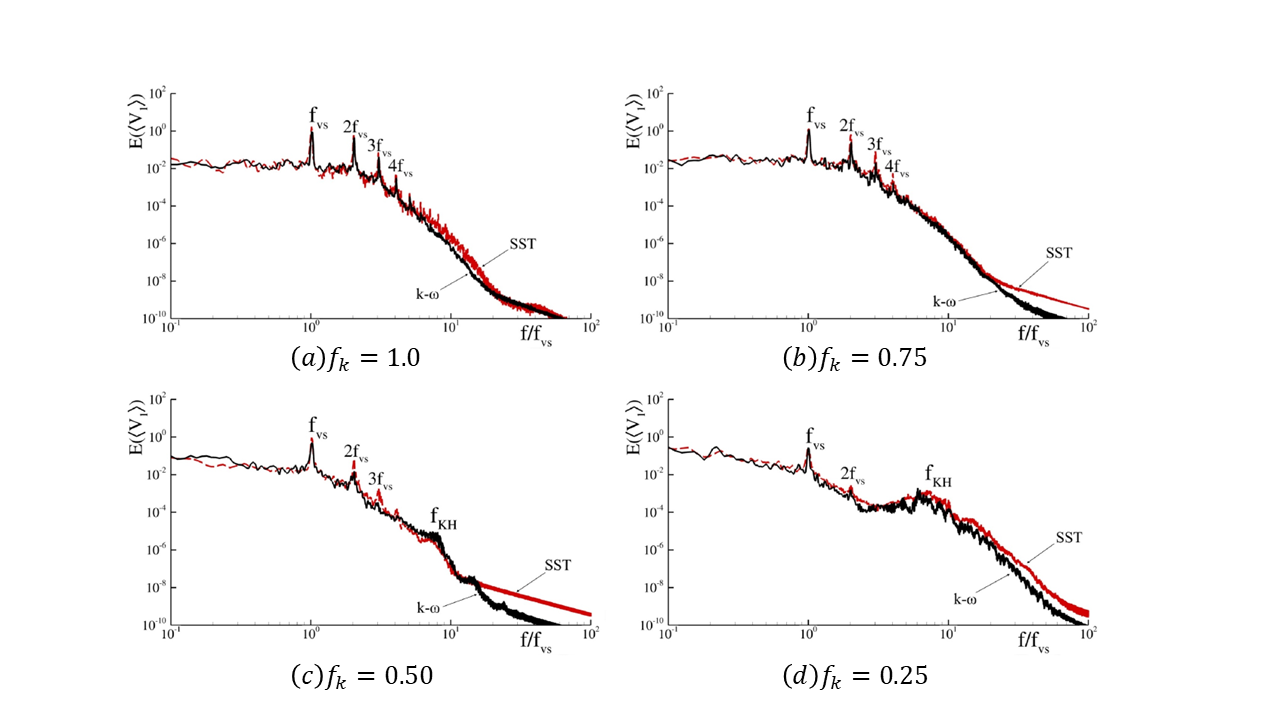}
    }
    \caption{\textbf{The mechanics of coherent structure emergence in cylinder-wake PANS simulations}. (a) Transition from over-saturated hot spots to local equilibrium; (b) $Re_{eff}$ crossing the critical Bloor threshold; (c) Spectral bifurcation where KH rollers emerge and compete with primary vortex shedding harmonics.}
    \label{fig:pans_comprehensive}
\end{figure*}

\paragraph*{Cases 1 \& 2: The Over-Saturated Branch ($f_k = 1.0, 0.75$).} 
At these coarse resolutions, the modeled wake is characterized by extreme shear \textit{hot spots} where $S K_u/\varepsilon_u > 10$ (Fig.~\ref{fig:pans_comprehensive}(a), panels a and b). This massive over-saturation ($\Pi \approx 9$) indicates that the unresolved bath is pushed far from equilibrium. In attempting to dissipate the imposed energy, the model produces unphysically high eddy viscosity, driving the effective computational Reynolds number well below the critical Bloor limit ($Re \approx 1200$) \cite{Bloor1964} in the inflection line region (Fig.~\ref{fig:pans_comprehensive}(b)). Because the simulated resolved flow is effectively sub-critical, the KH instability is artificially suppressed. The spectra (Fig.~\ref{fig:pans_comprehensive}c) provide the final proof of this state, revealing only the primary vortex shedding frequency ($f_{vs}$) and its first four harmonics. The simulated flow remains trapped on a lower-order thermodynamic branch, its structural complexity stifled by an over-saturated modeled bath.

\paragraph*{Case 3: The Thermodynamic Bifurcation Threshold ($f_k = 0.5$).} 
This resolution represents the critical phase transition. With the liberation of these structures, the dissipative tension is finally released along the inflection line as $S K_u/\varepsilon_u$ contours relax to the equilibrium range ($\approx 3.3$) (Fig.~\ref{fig:pans_comprehensive}(a), panel c). This relaxation signifies that the modeled bath has achieved local equilibrium ($\Pi \approx 1$), while the global system remains a Non-Equilibrium Steady State (NESS) sustained by macroscopic fluxes. Correspondingly, the eddy viscosity reduces, allowing the effective computational Reynolds number in this region to clear the Bloor threshold (Fig.~\ref{fig:pans_comprehensive}(b), panel c). The impact is immediate: KH rollers physically emerge, and a broadband \textit{hump} appears at the corresponding frequency ($f_{KH}$) in the spectrum (Fig.~\ref{fig:pans_comprehensive}(c), panel c). This distinct spectral bifurcation confirms that the resolved flow has successfully transitioned to the higher-order thermodynamic branch.

\paragraph*{Case 4: The Satisfied Mandate ($f_k = 0.25$).} 
At $f_k = 0.25$, the $S K_u/\varepsilon_u$ values drop globally below the equilibrium threshold (Fig.~\ref{fig:pans_comprehensive}(a), panel d), and $Re_{eff}$ is maintained well above the Bloor limit throughout the wake. The unresolved turbulent bath is now truly thermal, and the resolved field is free to fully accommodate the coherent structures. In the spectra (Fig.~\ref{fig:pans_comprehensive}(c), panel d), the KH rollers now dominate the secondary spectral energy, completely nullifying the higher-order VS harmonics. By shedding the mandated structures, the system successfully alleviates the localized throughput congestion.

As indicated by the amplitude equations of weakly nonlinear theory \cite{Cross1993}, these secondary bifurcations do not merely superimpose upon the primary state; they actively compete with it. In classical Rayleigh-B\'enard convection, this manifests as the saturation of primary 2D rolls, which are systematically suppressed and deformed by emergent 3D cross-roll and plume instabilities to accommodate elevated thermodynamic fluxes \cite{Ahlers2009}—a transition Prigogine fundamentally characterized as a cascading evolutionary sequence of dissipative structures \cite{NicolisPrigogine1977}.


\section{Synthesis: A Universal Blueprint for Dissipative Organization}
\label{sec:synthesis_blueprint}

The theoretical and operational frameworks developed in the preceding sections can now be synthesized into a unified physical narrative. We propose that the emergence of coherent structures in turbulence is not a kinematic accident, but a predictable sequence of events driven by throughput imbalance. This structural emergence defines the core engine of non-equilibrium turbulence and unfolds as a universal blueprint across three distinct phases:

\paragraph*{Phase I: The Linear Branch and Broadening of the Bath.}
The cycle begins with macroscopic external forcing (e.g., mean shear, pressure gradients) continually injecting energy into the resolved mean flow. To maintain statistical stationarity, the system attempts to remain on its unstructured, thermodynamic "linear branch" by irreversibly cascading this energy down to the unresolved scales, representing a localized unresolved production ($P_u$). As the imposed forcing increases, the system initially responds by simply broadening the spectrum of the structureless bath, increasing the unresolved dissipation ($\varepsilon_u$) to match the production without altering the macroscopic topology of the flow.

\paragraph*{Phase II: The Dissipative Crisis and Kinematic Criticality.}
As forcing continues to escalate, the unstructured microscopic bath eventually reaches its absolute dissipative capacity. The local throughput, quantified by the unresolved production-to-dissipation ratio $\Pi \equiv P_u / \varepsilon_u$, exceeds unity, indicating that unresolved production critically outpaces unresolved dissipation. 

We term this thermodynamic bottleneck a \emph{Dissipative Crisis}. The unstructured NESS branch can no longer accommodate the imposed energy flux through simple, isotropic energy transfer. To avoid unbounded energy accumulation and a catastrophic departure from stationarity, the thermodynamic framework strictly demands a structural reorganization of the flow.

\paragraph*{Phase III: Coherence Liberation and Nonlinear Saturation.}
To resolve the dissipative crisis, the fluid system leverages the dynamically admissible pathways of the fluctuation-moderated mean state ($\bar{\mathbf{U}}$). The linear NESS generator, $J(X_0)$, dictates how the flow can reorganize to relieve the energetic bottleneck, executing \emph{coherence liberation} through one of two primary mechanisms:
\begin{itemize}
    \item \textbf{Modal Branch Access (Wake Flows):} If the NESS generator possesses an unstable eigenvalue, the flow undergoes a classical Prigoginian bifurcation. The instability acts as the vehicle that transforms the system onto a new, structurally organized thermodynamic branch (e.g., vortex shedding in wakes).
    \item \textbf{Transient Relief / Non-Modal Growth (Shear Flows):} In modally stable flows, the highly non-normal Navier--Stokes operator permits massive transient amplification. The system leverages resolvent amplification to temporarily but robustly construct the organized motions required to bypass the subgrid bottleneck and accelerate global energy transfer (e.g., streaks and roll-cells in wall-bounded turbulence).
\end{itemize}
Once triggered, these amplified linear instabilities grow until they interact nonlinearly, saturating into a new, stable sequence of macroscopic coherent structures. These organized motions---the nonlinear attractors of the coarse-grained NESS---serve as the highly efficient, multiscale ``conveyor belts'' required to restore the global production--dissipation balance to the system.

\paragraph*{Coherent Structures in Open and Closed Turbulent Systems.}
The dual-regime framework introduced herein ultimately exposes a grand, unifying teleological principle governing open and closed turbulent systems: structure is invariably invoked by nature as a stabilizing thermodynamic defense mechanism to prevent a catastrophic local accumulation of energy.
 Every driven fluid system faces a fundamental crisis when the localized energy throughput events threaten to overwhelm its baseline relaxation pathways. The system resolves this imbalance by drastically accelerating its effective rate of dissipation, utilizing one of two distinct structural avenues depending on its thermodynamic boundaries. In statistically closed or unforced settings, exemplified by the classical frameworks of Onsager \cite{onsager1949}, Kraichnan \cite{Kraichnan1959}, and Eyink and Sreenivasan \cite{EyinkSreenivasan2006}, the crisis drives the formation of \emph{microscopic} small scales, with the mechanism supplied entirely by the down-scale inertial cascade. Conversely, in the open, continuously driven systems investigated here, the imbalance liberates \emph{macroscopic} coherent structures. Depending on the system's topological access to bifurcation, this liberation either manifests as transient, accommodative adjustments within the baseline branch (Class I) or as a transformative break away to entirely new, autonomous non-linear solution manifolds (Class II). Crucially, whether the system creates non-differentiable microscopic singularities or self-organizes into highly ordered macroscopic rollers and streaks, the underlying thermodynamic mandate is identical: to fundamentally enhance transport, accelerate energy transfer, and maximize effective dissipation to preserve the stable, bounded regulation of the flow physics.

\paragraph*{Practical Implications for Computational Physics.}
The theoretical reframing presented herein offers profound practical utility for computational physics, directly fulfilling the operational mandate established in the Introduction. It is widely recognized that non-equilibrium coherent structures must be explicitly resolved, whereas equilibrium stochastic turbulent motions can be reliably modeled. By identifying coherent structures as fundamentally dissipative phenomena, we establish a first-principles, thermodynamic criterion for Scale-Resolving Simulations (SRS). The order parameter $\Pi$ of the unresolved scales serves as a computable localized metric to assess the dissipative tension. 

In simulations of flows characterized by modal instabilities (bifurcations), the emergence of persistent dissipative tension dictates that the underlying structure must be explicitly resolved---typically by increasing resolution (e.g., reducing the modeled fraction $f_k$)---until $\Pi = P_u/\varepsilon_u$ is restored to the order of unity. This targeted resolution effectively alleviates excessive subgrid damping, liberates the macroscopic flow structures, and relieves the dissipative bottleneck. Consequently, the predictive simulation is constrained to the exact thermodynamic stability branch as the physical flow, guaranteeing physically reliable results. 

Conversely, in simulations of modally stable flows, the order parameter $\Pi$ remains intrinsically close to unity. In these regimes, there are no abrupt structural bifurcations tied to specific resolution thresholds. Rather, as demonstrated by Kamble et al.\ \cite{Kambleetal2022}, the fidelity of the flow prediction improves continuously and smoothly with increasing resolution (decreasing $f_k$ or decreasing computational Reynolds number). 

Thus, the characterization of coherence as dissipative organization directly yields actionable metrics for practical computational fluid dynamics. Consequently, the accurate computation of coherent structures is elevated from an exercise in morphological curve-fitting to the mandatory resolution of the fluid system's underlying thermodynamic degrees of freedom.


\section{Conclusion and Future Outlook}
\label{sec:conclusion_outlook}

In this work, we establish that turbulent coherent structures are not merely morphological features, but fundamental thermodynamic entities---specifically, \emph{generalized dissipative structures}. Their emergence is strictly mandated by a throughput imbalance within an open, coarse-grained Navier--Stokes (CGNS) bath. This framework mathematically unifies three critical aspects of turbulence:
\begin{enumerate}[leftmargin=*]
    \item \textbf{Modal and Non-Modal Structures:} Synthesizing turbulent structures with Prigogine's far-from-equilibrium and Kubo's near-equilibrium thermodynamic exemplars. 
    \item \textbf{Open and Closed Systems:} Bridging coherent structures (present work) and the turbulent cascade (\cite{EyinkSreenivasan2006}) under a unified thermodynamic imperative: structural emergence is a mandatory physical response to avert macroscopic energetic runaway.
    \item \textbf{Diagnostic Techniques:} Reinterpreting disparate morphology-based methods (e.g., POD, DMD, LCS) as complementary, physically grounded projections of a single underlying thermodynamic reality.
\end{enumerate}

This formulation yields a universal, representation-independent criterion for structural emergence: the order parameter $\Pi = P_u/\varepsilon_u$. By transitioning resolution criteria from empirical calibration to fundamental thermodynamics, $\Pi$ serves not merely as an \textit{a posteriori} diagnostic, but as an active physical constraint for dynamically adapting scale-resolving simulations.

While experimental investigations of turbulence have historically excelled in unraveling the intricate details of dynamical processes, there has been less focus on identifying a quantity that captures the underlying organizational mandate of turbulent flows. The thermodynamic order parameter introduced in this work, $\Pi = P_u / \varepsilon_u$, provides a unified perspective, opening a new direction for experimental inquiry. This framework does not supplant prior achievements but offers a complementary lens: it empowers future experiments to directly probe the fundamental throughput imbalances that drive the emergence of organization. Consequently, this theory guides experimental focus beyond isolated dynamical events, pointing toward the universal thermodynamic imperatives that govern structural self-organization.

Beyond physical experiments, this approach establishes a rigorous basis for computational structure identification. By casting modeled CGNS equations as a direct operationalization of this statistical mechanics framework, we transition predictive flow modeling from phenomenological closure to rigorous, first-principles computation. Ultimately, this paradigm recasts complex, multiscale fluid dynamics strictly through the universal thermodynamic lens of macroscopic forcing and mesoscale entropy production.

\section*{Acknowledgments}
The support provided by the ABS Department Head chair is gratefully acknowledged. The author would like to thank Dr. Filipe Pereira for inspiring this work. The author additionally acknowledges the use of artificial intelligence tools (specifically, Google Gemini) to assist with language editing, and structural formatting. The author has reviewed all generated output and takes full responsibility for the scientific content and integrity of this work.

\paragraph*{Funding:}
This work was not supported by any external funding.

\paragraph*{Author contributions:}
S.S.G. is the sole author of this paper and was responsible for all conceptualization, methodology, analysis, and writing.

\paragraph*{Competing interests:}
The author declares that he has no competing interests.

\paragraph*{Data and materials availability:}
All data needed to evaluate the conclusions in the paper are present in the paper. The primary computational datasets analyzed in this study are publicly available in the referenced articles.


\clearpage 

\bibliography{science_template} 

@book{feynman1964lectures,
  title={The Feynman Lectures on Physics, Vol. II: Mainly Electromagnetism and Matter},
  author={Feynman, Richard P. and Leighton, Robert B. and Sands, Matthew},
  year={1964},
  publisher={Addison-Wesley},
  address={Reading, MA}
}

@article{Kolmogorov1941,
  title={The local structure of turbulence in incompressible viscous fluid for very large {R}eynolds numbers},
  author={Kolmogorov, Andrey Nikolaevich},
  journal={Proc. R. Soc. A},
  volume={434},
  number={1890},
  pages={9--13},
  year={1991},
  publisher={The Royal Society London},
  note={Original Russian publication in Dokl. Akad. Nauk SSSR 30, 299--303 (1941)}
}

@article{Kolmogorov1962,
  title={A refinement of previous hypotheses concerning the local structure of turbulence in a viscous incompressible fluid at high {R}eynolds number},
  author={Kolmogorov, Andrey Nikolaevich},
  journal={J. Fluid Mech.},
  volume={13},
  number={1},
  pages={82--85},
  year={1962},
  publisher={Cambridge University Press}
}

@book{Batchelor1953,
  author    = {Batchelor, G. K.},
  title     = {The Theory of Homogeneous Turbulence},
  publisher = {Cambridge University Press},
  year      = {1953}
}

@book{Frisch1995,
  author    = {Frisch, U.},
  title     = {Turbulence: The Legacy of {A}. {N}. {K}olmogorov},
  publisher = {Cambridge University Press},
  address   = {Cambridge},
  year      = {1995}
}

@article{Germano1992,
  author  = {Germano, M.},
  title   = {Turbulence: The filtering approach},
  journal = {J. Fluid Mech.},
  volume  = {238},
  pages   = {325--336},
  year    = {1992},
  doi     = {10.1017/S0022112092001733}
}

@article{Girimaji2006,
  author  = {Girimaji, S. S.},
  title   = {Partially-averaged {N}avier--{S}tokes model for turbulence: A {R}eynolds-averaged {N}avier--{S}tokes to direct numerical simulation bridging method},
  journal = {J. Fluids Eng.},
  volume  = {128},
  pages   = {413--421},
  year    = {2006},
  doi     = {10.1115/1.2151207}
}

@article{Girimaji2024,
  author  = {Girimaji, S. S.},
  title   = {Turbulence closure modeling with machine learning: A foundational physics perspective},
  journal = {New J. Phys.},
  volume  = {26},
  number  = {7},
  pages   = {071201},
  year    = {2024},
  doi     = {10.1088/1367-2630/ad6689}
}

@article{Hussain1986,
  author  = {Hussain, A. K. M. F.},
  title   = {Coherent structures and turbulence},
  journal = {J. Fluid Mech.},
  volume  = {173},
  pages   = {303--356},
  year    = {1986},
  doi     = {10.1017/S0022112086001192}
}

@incollection{Lumley1967,
  author    = {Lumley, J. L.},
  title     = {The structure of inhomogeneous turbulent flows},
  booktitle = {Atmospheric Turbulence and Radio Wave Propagation},
  pages     = {166--178},
  year      = {1967}
}

@inproceedings{Lumley1981_CoherentStructures,
  author    = {Lumley, J. L.},
  title     = {Coherent structures in turbulence},
  booktitle = {Transition and Turbulence},
  editor    = {Meyer, R. E.},
  pages     = {215--242},
  publisher = {Academic Press},
  year      = {1981}
}

@article{McKeonSharma2010,
  author  = {McKeon, B. J. and Sharma, A. S.},
  title   = {A critical-layer framework for turbulent pipe flow},
  journal = {J. Fluid Mech.},
  volume  = {658},
  pages   = {336--382},
  year    = {2010},
  doi     = {10.1017/S002211201000176X}
}

@book{MoninYaglom1975,
  author    = {Monin, A. S. and Yaglom, A. M.},
  title     = {Statistical Fluid Mechanics, Volume 2},
  publisher = {MIT Press},
  address   = {Cambridge, MA},
  year      = {1975}
}

@book{NicolisPrigogine1977,
  author    = {Nicolis, G. and Prigogine, I.},
  title     = {Self-Organization in Nonequilibrium Systems},
  publisher = {Wiley},
  address   = {New York},
  year      = {1977}
}

@article{Onsager1931_I,
  author  = {Onsager, L.},
  title   = {Reciprocal relations in irreversible processes. {I}.},
  journal = {Phys. Rev.},
  volume  = {37},
  number  = {4},
  pages   = {405--426},
  year    = {1931},
  doi     = {10.1103/PhysRev.37.405}
}

@book{Prigogine1967,
  author    = {Prigogine, I.},
  title     = {Introduction to Thermodynamics of Irreversible Processes},
  publisher = {Wiley},
  address   = {New York},
  year      = {1967}
}

@article{Schmid2010,
  author  = {Schmid, P. J.},
  title   = {Dynamic mode decomposition of numerical and experimental data},
  journal = {J. Fluid Mech.},
  volume  = {656},
  pages   = {5--28},
  year    = {2010}
}

@article{Sirovich1987_CoherentStructures,
  author  = {Sirovich, L.},
  title   = {Turbulence and the dynamics of coherent structures. {P}art {I}: Coherent structures},
  journal = {Q. Appl. Math.},
  volume  = {45},
  number  = {3},
  pages   = {561--571},
  year    = {1987},
  doi     = {10.1090/qam/910462}
}

@misc{Prigogine1977Nobel,
  author       = {Prigogine, I.},
  title        = {Time, structure and fluctuations},
  note         = {Nobel lecture, December 8, 1977},
  year         = {1977},
  howpublished = {\url{https://www.nobelprize.org/prizes/chemistry/1977/prigogine/lecture/}}
}

@article{Ahlers2009,
  author  = {Ahlers, G. and Grossmann, S. and Lohse, D.},
  title   = {Heat transfer and large scale dynamics in turbulent {R}ayleigh--{B}{\'e}nard convection},
  journal = {Rev. Mod. Phys.},
  volume  = {81},
  number  = {2},
  pages   = {503--537},
  year    = {2009},
  doi     = {10.1103/RevModPhys.81.503}
}

@article{Esposito2012,
  author  = {Esposito, M.},
  title   = {Stochastic thermodynamics under coarse graining},
  journal = {Phys. Rev. E},
  volume  = {85},
  number  = {4},
  pages   = {041125},
  year    = {2012},
  doi     = {10.1103/PhysRevE.85.041125}
}

@article{Cross1993,
  author  = {Cross, M. C. and Hohenberg, P. C.},
  title   = {Pattern formation outside of equilibrium},
  journal = {Rev. Mod. Phys.},
  volume  = {65},
  number  = {3},
  pages   = {851--1112},
  year    = {1993},
  doi     = {10.1103/RevModPhys.65.851}
}

@book{Goldenfeld1992,
  author    = {Goldenfeld, N.},
  title     = {Lectures on Phase Transitions and the Renormalization Group},
  publisher = {CRC Press},
  address   = {Boca Raton, FL},
  year      = {1992}
}

@book{Cercignani1988,
  author    = {Cercignani, C.},
  title     = {The {B}oltzmann Equation and Its Applications},
  publisher = {Springer-Verlag},
  address   = {New York},
  year      = {1988}
}

@article{onsager1949,
  author  = {Onsager, L.},
  title   = {Statistical hydrodynamics},
  journal = {Nuovo Cimento Suppl.},
  volume  = {6},
  pages   = {279--287},
  year    = {1949},
  doi     = {10.1007/BF02780991}
}

@book{Glansdorff1971,
  author    = {Glansdorff, P. and Prigogine, I.},
  title     = {Thermodynamic Theory of Structure, Stability and Fluctuations},
  publisher = {Wiley-Interscience},
  address   = {London},
  year      = {1971},
  isbn      = {978-0471302803}
}

@book{nicolis1977,
  author    = {Nicolis, G. and Prigogine, I.},
  title     = {Self-Organization in Nonequilibrium Systems},
  publisher = {Wiley},
  address   = {New York},
  year      = {1977}
}

@article{Girimaji2024_NJP,
  author  = {Girimaji, S. S.},
  title   = {Turbulence closure modeling with machine learning: A foundational physics perspective},
  journal = {New J. Phys.},
  volume  = {26},
  number  = {7},
  pages   = {071201},
  year    = {2024},
  doi     = {10.1088/1367-2630/ad6689}
}

@book{Townsend1956,
  title={The Structure of Turbulent Shear Flow},
  author={Townsend, Albert Alan},
  year={1956},
  publisher={Cambridge University Press},
  address={Cambridge, UK}
}

@article{Pereira2018JCP,
  author  = {Pereira, Filipe S. and E{\c{c}}a, Lu{\'i}s and Vaz, Guilherme and Girimaji, Sharath S.},
  title   = {Challenges in Scale-Resolving Simulations of turbulent wake flows with coherent structures},
  journal = {J. Comput. Phys.},
  volume  = {363},
  pages   = {98--115},
  year    = {2018},
  doi     = {10.1016/j.jcp.2018.02.038}
}

@article{Hamilton1995,
  title={Regeneration mechanisms of near-wall turbulence structures},
  author={Hamilton, J. M. and Kim, J. and Waleffe, F.},
  journal={J. Fluid Mech.},
  volume={287},
  pages={317--348},
  year={1995},
  publisher={Cambridge University Press}
}

@article{Waleffe1997,
  title={On a self-sustaining process in shear flows},
  author={Waleffe, F.},
  journal={Phys. Fluids},
  volume={9},
  number={4},
  pages={883--900},
  year={1997},
  publisher={American Institute of Physics}
}

@article{Wallace1972,
  title={The wall region in turbulent shear flow},
  author={Wallace, J. M. and Eckelmann, H. and Brodkey, R. S.},
  journal={J. Fluid Mech.},
  volume={54},
  number={1},
  pages={39--48},
  year={1972},
  publisher={Cambridge University Press}
}

@article{Willmarth1972,
  title={Structure of the Reynolds stress near the wall},
  author={Willmarth, W. W. and Lu, S. S.},
  journal={J. Fluid Mech.},
  volume={55},
  number={1},
  pages={65--92},
  year={1972},
  publisher={Cambridge University Press}
}

@article{Jimenez1991,
  title={The minimal flow unit in near-wall turbulence},
  author={Jim{\'e}nez, J. and Moin, P.},
  journal={J. Fluid Mech.},
  volume={225},
  pages={213--240},
  year={1991},
  publisher={Cambridge University Press}
}

@article{Williamson1996,
  title={Vortex dynamics in the cylinder wake},
  author={Williamson, C. H. K.},
  journal={Annu. Rev. Fluid Mech.},
  volume={28},
  pages={477--539},
  year={1996},
  publisher={Annual Reviews}
}

@article{BarkleyHenderson1996,
  title={Three-dimensional Floquet stability analysis of the wake of a circular cylinder},
  author={Barkley, D. and Henderson, R. D.},
  journal={J. Fluid Mech.},
  volume={322},
  pages={215--241},
  year={1996},
  publisher={Cambridge University Press}
}

@article{Bloor1964,
  title={The transition to turbulence in the wake of a circular cylinder},
  author={Bloor, M. S.},
  journal={J. Fluid Mech.},
  volume={19},
  pages={290--304},
  year={1964},
  publisher={Cambridge University Press}
}

@techreport{Norberg1987,
  title={Effects of Reynolds number and free-stream turbulence on the flow around a circular cylinder},
  author={Norberg, C.},
  institution={Chalmers University of Technology},
  address={G{\"o}teborg, Sweden},
  year={1987},
  number={87/2}
}

@article{Trefethen1993,
  title={Hydrodynamic stability without eigenvalues},
  author={Trefethen, Lloyd N and Trefethen, Anne E and Reddy, Satish C and Driscoll, Tobin A},
  journal={Science},
  volume={261},
  number={5121},
  pages={578--584},
  year={1993},
  publisher={American Association for the Advancement of Science},
  doi={10.1126/science.261.5121.578}
}

@article{Kubo1966,
  title={The fluctuation-dissipation theorem},
  author={Kubo, Ryogo},
  journal={Rep. Prog. Phys.},
  volume={29},
  number={1},
  pages={255--284},
  year={1966},
  publisher={IOP Publishing}
}

@article{Farrell1993,
  title={Stochastic forcing of the linearized Navier--Stokes equations},
  author={Farrell, Brian F and Ioannou, Petros J},
  journal={Phys. Fluids A},
  volume={5},
  number={11},
  pages={2600--2609},
  year={1993},
  publisher={American Institute of Physics}
}

@article{EyinkSreenivasan2006,
  title={Onsager and the theory of hydrodynamic turbulence},
  author={Eyink, Gregory L and Sreenivasan, Katepalli R},
  journal={Rev. Mod. Phys.},
  volume={78},
  number={1},
  pages={87--135},
  year={2006},
  publisher={American Physical Society},
  doi={10.1103/RevModPhys.78.87}
}

@book{KuboTodaHashitsume1991,
  author    = {Kubo, R. and Toda, M. and Hashitsume, N.},
  title     = {Statistical Physics II: Nonequilibrium Statistical Mechanics},
  publisher = {Springer},
  year      = {1991}
}

@article{Taghizadehetal2020,
  author  = {Taghizadeh, Salar and Witherden, Freddie D. and Girimaji, Sharath S.},
  title   = {Turbulence closure modeling with data-driven techniques: physical compatibility and consistency considerations},
  journal = {New J. Phys.},
  year    = {2020},
  volume  = {22},
  number  = {9},
  pages   = {093023},
  doi     = {10.1088/1367-2630/abadb3}
}

@article{TazraeiGirimaji2019,
  title = {Scale-resolving simulations of turbulence: Equilibrium boundary layer analysis leading to near-wall closure modeling},
  author = {Tazraei, Pedram and Girimaji, Sharath S.},
  journal = {Phys. Rev. Fluids},
  volume = {4},
  issue = {10},
  pages = {104607},
  numpages = {25},
  year = {2019},
  month = {Oct},
  publisher = {American Physical Society},
  doi = {10.1103/PhysRevFluids.4.104607},
  url = {https://link.aps.org/doi/10.1103/PhysRevFluids.4.104607}
}

@article{Kambleetal2022,
  title = {Closure modeling in near-wall region of steep resolution variation for partially averaged Navier-Stokes simulations},
  author = {Kamble, Chetna and Girimaji, Sharath and Razi, Pooyan and Tazraei, Pedram and Wallin, Stefan},
  journal = {Phys. Rev. Fluids},
  volume = {7},
  issue = {4},
  pages = {044608},
  numpages = {25},
  year = {2022},
  month = {Apr},
  publisher = {American Physical Society},
  doi = {10.1103/PhysRevFluids.7.044608},
  url = {https://link.aps.org/doi/10.1103/PhysRevFluids.7.044608}
}

@article{Bertini2015,
  author  = {L. Bertini and A. De Sole and D. Gabrielli and G. Jona-Lasinio and C. Landim},
  title   = {Macroscopic fluctuation theory},
  journal = {Rev. Mod. Phys.},
  volume  = {87},
  number  = {2},
  pages   = {593--636},
  year    = {2015}
}

@article{kim1987turbulence,
    author  = {Kim, J. and Moin, P. and Moser, R.},
    title   = {Turbulence statistics in fully developed channel flow at low Reynolds number},
    journal = {J. Fluid Mech.},
     volume  = {177},
     pages   = {133--166},
      year    = {1987}
}

@article{moser1999direct,
     author  = {Moser, R. and Kim, J. and Mansour, N.},
    title   = {Direct numerical simulation of turbulent channel flow},
    journal = {Phys. Fluids},
    volume  = {11},
     number  = {4},
    pages   = {943--945},
    year    = {1999}
}

@article{schlatter2010assessment,
   author  = {Schlatter, P. and \"Orl\"u, R.},
   title   = {Assessment of direct numerical simulation data of turbulent boundary layers},
   journal = {J. Fluid Mech.},
   volume  = {659},
   pages   = {116--126},
   year    = {2010}
}

@article{lozano2014effect,
   author  = {Lozano-Dur{\'a}n, A. and Jim{\'e}nez, J.},
   title   = {Effect of the computational domain on direct simulations of turbulent channels},
   journal = {Phys. Fluids},
   volume  = {26},
   number  = {1},
   pages   = {011702},
   year    = {2014}
}

@incollection{anderson1984,
  author    = {Anderson, P. W. and Stein, D. L.},
  title     = {Broken Symmetry, Emergent Properties, Dissipative Structures, Life: Are They Related?},
  booktitle = {Self-Organizing Systems},
  publisher = {Springer},
  address   = {Boston, MA},
  pages     = {445--465},
  year      = {1984}
}

@article{bricmont1995,
  author  = {Bricmont, J.},
  title   = {Science of Chaos or Chaos in Science?},
  journal = {Physicalia},
  volume  = {17},
  pages   = {159--208},
  year    = {1995}
}

@article{girimaji2007,
  author    = {Girimaji, S. S.},
  title     = {Boltzmann Kinetic Equation for Filtered Fluid Turbulence},
  journal   = {Phys. Rev. Lett.},
  volume    = {99},
  issue     = {3},
  pages     = {034501},
  year      = {2007},
  publisher = {American Physical Society}
}

@article{OnsagerMachlup1953,
  title = {Fluctuations and Irreversible Processes},
  author = {Onsager, L. and Machlup, S.},
  journal = {Phys. Rev.},
  volume = {91},
  issue = {6},
  pages = {1505--1512},
  numpages = {0},
  year = {1953},
  month = {Sep},
  publisher = {American Physical Society},
  doi = {10.1103/PhysRev.91.1505},
  url = {https://link.aps.org/doi/10.1103/PhysRev.91.1505}
}

@article{TazraeiGirimaji2020,
  title = {Scale-resolving simulations of spatially evolving turbulence: Physically consistent inflow specification of unresolved velocity and length-scale profiles},
  author = {Tazraei, Pedram and Girimaji, Sharath S.},
  journal = {Phys. Rev. Fluids},
  volume = {5},
  issue = {12},
  pages = {124604},
  numpages = {26},
  year = {2020},
  month = {Dec},
  publisher = {American Physical Society},
  doi = {10.1103/PhysRevFluids.5.124604},
  url = {https://link.aps.org/doi/10.1103/PhysRevFluids.5.124604}
}

@article{Kraichnan1959,
  author    = {Kraichnan, Robert H.},
  title     = {The structure of isotropic turbulence at very high {R}eynolds numbers},
  journal   = {J. Fluid Mech.},
  volume    = {5},
  number    = {4},
  pages     = {497--543},
  year      = {1959},
  publisher = {Cambridge University Press},
  doi       = {10.1017/S0022112059000362}
}

@article{Berge1976,
  title = {Experimental Confirmation of Fluctuation Amplification Near the Rayleigh-B\'enard Instability},
  author = {Berg\'e, P. and Dubois, M.},
  journal = {Phys. Rev. Lett.},
  volume = {36},
  issue = {18},
  pages = {1083--1086},
  numpages = {4},
  year = {1976},
  month = {May},
  publisher = {American Physical Society},
  doi = {10.1103/PhysRevLett.36.1083},
  url = {https://link.aps.org/doi/10.1103/PhysRevLett.36.1083}
}

@article{Zaitsev1971,
  title = {Hydrodynamic Fluctuations Near the Convection Threshold},
  author = {Zaitsev, V. M. and Shliomis, M. I.},
  journal = {Sov. Phys. JETP},
  volume = {32},
  number = {5},
  pages = {866--870},
  year = {1971}
}

@article{Lekkerkerker1974,
  title = {Hydrodynamic fluctuations near the {Rayleigh-B\'enard} instability},
  author = {Lekkerkerker, H. N. W. and Boon, J. P.},
  journal = {Physica},
  volume = {73},
  number = {3},
  pages = {571--584},
  year = {1974},
  doi = {10.1016/0031-8914(74)90111-6}
}

@article{Graham1974,
  title = {Hydrodynamic fluctuations near the {Rayleigh-B\'enard} instability},
  author = {Graham, R.},
  journal = {Phys. Rev. A},
  volume = {10},
  number = {5},
  pages = {1762--1772},
  year = {1974},
  publisher = {American Physical Society},
  doi = {10.1103/PhysRevA.10.1762}
}

@article{Zaitsev1970tc,
  title = {Hydrodynamic Fluctuations Near the Couette Instability Threshold},
  author = {Zaitsev, V. M. and Shliomis, M. I.},
  journal = {Sov. Phys. JETP},
  volume = {32},
  number = {5},
  pages = {866--870},
  year = {1971}
}

@article{Snyder1970,
  title = {Fluctuations and the transition to {Taylor} vortex flow},
  author = {Snyder, H. A.},
  journal = {Phys. Fluids},
  volume = {13},
  number = {10},
  pages = {2437--2444},
  year = {1970},
  doi = {10.1063/1.1692813}
}

@article{Vidal1982,
  title = {Pretransitional fluctuations in a chemical system far from equilibrium},
  author = {Vidal, C. and Bachelart, A. and Rossi, A.},
  journal = {J. Phys. Chem.},
  volume = {86},
  number = {26},
  pages = {5163--5169},
  year = {1982},
  doi = {10.1021/j100223a021}
}

@article{Kai1978,
  title = {Hydrodynamic Fluctuation as a Precursor of {Williams} Domain Instability},
  author = {Kai, S. and Hirakawa, K.},
  journal = {J. Phys. Soc. Jpn.},
  volume = {44},
  number = {2},
  pages = {711--712},
  year = {1978},
  doi = {10.1143/JPSJ.44.711}
}
\bibliographystyle{sciencemag}

\end{document}